\documentclass[aps,prb,twocolumn,showpacs,superscriptaddress, showkeys]{revtex4-2}
\usepackage{graphicx}
\usepackage[caption=false]{subfig}
\usepackage{hyperref}
\hypersetup{
	colorlinks=true,
	linkcolor=blue,      
	urlcolor=blue,
	bookmarks=true,
	citecolor=blue}
\usepackage{siunitx}
\usepackage{amsmath}
\newcommand{\mrg}{\mbox{Mn$_2$Ru$_x$Ga}}
\newcommand{\tcmp}{T_{\text{comp}}}
\graphicspath{ {./img/} }
\usepackage{amssymb}

\begin{document}

\title{Quasi-static magnetization dynamics in a compensated ferrimagnetic half-metal - \mrg}
\author{Ajay~Jha}
\email{ajha@tcd.ie}
\author{Simon~Lenne}
\author{Gwena\"{e}l~Atcheson}
\author{Karsten~Rode}
\author{ J.M.D.\,Coey}
\author{Plamen~Stamenov}
\affiliation{CRANN, AMBER, and School of Physics, Trinity College Dublin, Dublin 2, Ireland}


\begin{abstract}
{\setlength{\parindent}{0cm}
Exploring anisotropy and diverse magnetization dynamics in specimens with vanishing magnetic moments presents a significant challenge using traditional magnetometry, as the low resolution of existing techniques hinders the ability to obtain accurate results. In this study, we delve deeper into the examination of magnetic anisotropy and quasi-static magnetization dynamics in \mrg\,(MRG) thin films, as an example of a compensated ferrimagnetic half-metal, by employing anomalous Hall effect measurements within a tetragonal crystal lattice system. Our research proposes an innovative approach to accurately determine the complete set of anisotropy constants of these MRG thin films.
To achieve this, we perform anomalous Hall voltage curve fitting, using torque models under the macrospin approximation, which allow us to obtain out-of-plane anisotropy constants $K_1=4.0\times10^4$ J m$^{-3}$ ($K_1/M=0.655$\,T) and $K_2=2.54\times10^4$ J m$^{-3}$ ($K_2/M=0.416$\,T), along with a weaker in-plane anisotropy constant $K_3=3.48\times10^3$ J m$^{-3}$ ($K_3/M=0.057$\,T). By additionally employing first-order reversal curves (FORC) and classical Preisach hysteresis (hysterons) models, we are able to validate the efficacy of the macrospin model in capturing the magnetic behavior of MRG thin films.
Furthermore, our investigation substantiates that the complex quasi-static magnetization dynamics of MRG thin films can be effectively modelled using a combination of hysteronic and torque models. This approach facilitates the exploration of both linear and non-linear quasi-static magnetization dynamics, in the presence of external magnetic field and/or current-induced effective fields, generated by the spin-orbit torque and spin transfer torque mechanisms. The detailed understanding of the quasi-static magnetization dynamics is a key prerequisite for the exploitation of in-phase and out-of-phase resonance modes in this material class, for high-bandwidth modulators/de-modulators, filters and oscillators for the high-GHz and low-THz frequency bands.
}
\end{abstract}
%
%
\maketitle
\section{Introduction}
\label{sec:introduction}
Spintronics-based devices have emerged as  highly-promising candidates for next-generation telecommunication applications due to their potential for efficient control of magnetic moments via electrical methods, ultrafast operating speeds, and ultralow power dissipation~\cite{chappert2007emergence}. In the pursuit of these capabilities, antiferromagnetic (AFM) materials have demonstrated exceptional advantages over their ferromagnetic (FM) counterparts when incorporated into spintronic devices~\cite{jungwirth2016antiferromagnetic,RevModPhys.90.015005}. FM materials are limited by their large stray field interactions and slow switching speeds (on the order of nanoseconds)~\cite{khvalkovskiy2013basic,demidov2012magnetic}, which hampers their utility in memory and switching devices. In contrast, AFM materials project no stray fields and possess ultrafast spin dynamics (on the order of a few picoseconds to hundreds of picoseconds)~\cite{wadley2016electrical}, making them attractive for spintronics applications. However, the control and detection of magnetization in AFM materials remain challenging due to their zero magnetic moment and zero Fermi level spin polarization.

To address this technological gap, compensated ferrimagnetic half-metal (CFHM)~\cite{PhysRevB.57.10613,PhysRevB.75.172405} materials have emerged as an excellent alternative to AFM materials. Similar to AFM materials, CFHMs also consist of two antiferromagnetically coupled spin sublattices, and their spin contribution can be conveniently tuned by adjusting the composition and/or temperature. Moreover, due to the presence of inequivalent spin sites, these sublattices contribute unequally at the Fermi level, resulting in a semiconducting band-gap in one of the spin channels and a zero band gap in the other~\cite{Kurt2014}. This disparity gives rise to a high spin polarization for the conduction electrons.

At the magnetic compensation point, CFHMs exhibit behaviour akin to AFM materials, demonstrating ultrafast magnetization dynamics. However, unlike AFM materials, the detection and manipulation of CFHM magnetization remain feasible due to the distinct responses of the two spin sublattices to electrical and optical excitation~\cite{Siewierska2021,banerjee2020single,teichert2021magnetic}. These unique characteristics of CFHM pave the way for the development of novel spintronics devices with improved performance and functionality.

\begin{figure}[htb]
\centering
    \includegraphics[width=0.80\linewidth]{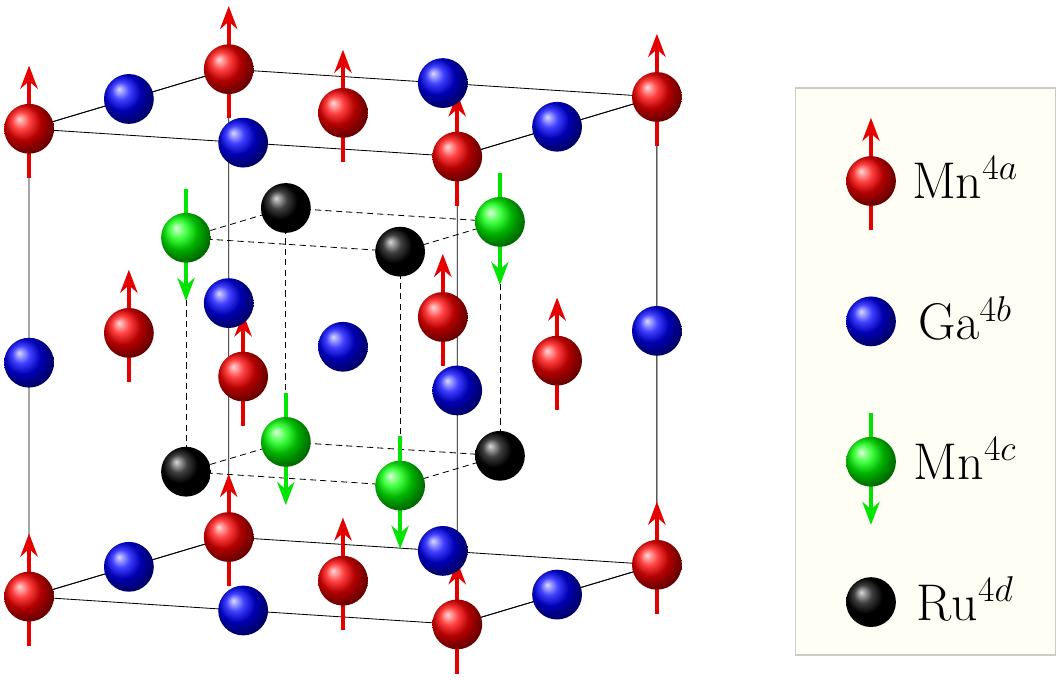}
    \captionsetup{singlelinecheck=false,labelformat=simple, width=\linewidth}
    \caption{\label{fig:crystal2} Schematic representation of the crystal structure of the MRG compound (Mn$_2$RuGa). The structure features four interpenetrating face-centered cubic lattices, with Mn$^{4a}$ (red) atoms occupying the $4a$ Wyckoff positions and Mn$^{4c}$ (green) atoms at the $4c$ positions. Ru (black) and Ga (blue) atoms are situated at the 4d and 4b Wyckoff positions, respectively. The magnetic moments at the Mn$^{4a}$ and Mn$^{4c}$ sites exhibit antiferromagnetic coupling. In this depiction, both the $4c$ and $4d$ positions are fully occupied by Mn and Ru atoms, respectively.}
    \vspace{-0.5cm}
\end{figure}

Following theoretical predictions~\cite{van1995half}, the first experimental observation of a CFHM thin film was achieved with the  \mrg\,(MRG) class of materials~\cite{Kurt2014}.  MRG crystallises in the inverse-Heusler XA structure, space group $ F\bar{4}3m $,  as depicted in figure~\ref{fig:crystal2}. Within this structure, Mn occupies two distinct and non-equivalent sublattice sites: $ 4a $ (Mn$^{4a} $) and $ 4c $  (Mn$^{4c} $). 
The Magnetic moments of $ 4a $ and $ 4c $ sublattices  exhibit antiferromagnetic coupling, whereas moments located on identical sites display ferromagnetic coupling.  Owing to the non-equivalent crystallographic surroundings, the magnetic moments of both sublattices display quite distinct temperature-dependent characteristics. The site-specific magnetic moment attributable to Mn$^{4a} $ possesses weaker temperature dependence in contrast to  Mn$^{4c} $ sublattice moment~\cite{betto2015site}. As such, it is possible to attain an ideal magnetic compensation for MRG by modulating its composition and/or inducing crystal lattice distortion.

MRG displays pronounced $c$-axis magnetic anisotropy due to the evanescence of magnetic moments, with an anisotropy field surpassing 14 T in proximity to the compensation point.  Furthermore, the Fermi level of MRG is predominantly influenced by the electronic states originating from Mn situated at the $ 4c $ position, subsequently dictating the transport phenomena via Mn$^{4c} $ electrons~\cite{Fowley2018}.  Additionally, MRG displays half-metallic properties, as corroborated by Density Functional Theory (DFT) calculations~\cite{Zic2016} and Point Contact Andreev Reflection (PCAR) spectroscopy  (spin polarization obtained as high as $ P>60 \, \% $)~\cite{Kurt2014,Siewierska2021}. The highly spin polarized carriers lead to a large anomalous Hall effect (AHE)~\cite{Fowley2018} and magneto-optic Kerr effect (MOKE)~\cite{siewierska2018imaging,banerjee2020single} even at the perfect magnetic compensation.
Therefore, the distinctive amalgamation of a vanishing net magnetic moment, high spin polarization at the Fermi level, and high magnetic anisotropy designates MRG as a promising contender material for active layers of next-generation spintronics devices.

By employing MRG as an active layer in the spin-oscillator, sub-THz chip-to-chip communication could be achieved, as its spin excitations were found to reside in the necessary terahertz gap~\cite{troncoso2019}.  The sub-THz excitations of MRG were ascribed to its low magnetic moment, high uniaxial anisotropy field, and low Gilbert damping~\cite{troncoso2019,Awari2016}.  In addition, the tunability of the anisotropy constant and the moment in MRG afford the flexibility to modify the resonance frequencies of oscillators constructed with MRG. Consequently, determining the anisotropy constants of MRG thin-films is a crucial preliminary step in examining their magnetization dynamics under the influence of external stimuli.

Investigating anisotropy and other magnetization dynamics in a sample with a negligible magnetic moment is unattainable using conventional magnetometry techniques (VSM, SQUID, etc.) due to insufficient resolution and sensitivity. Furthermore, for a sample with an extremely small magnetic moment ($M$), both the anisotropy field ($H_a=2K/M$) and coercive field ($H_c$) typically diverge, rendering the measurement of magnetic anisotropy unfeasible with exceedingly large magnetic fields ($\mu_0 H>14$~T)~\cite{Fowley2018}.  Generally, anisotropy is assessed by applying an external magnetic field at a specific angle ($\theta_H$) to the magnetic easy axis and monitoring the corresponding changes in physical properties such as magnetization~\cite{fan2007approach,endo2000determination,jagla2005hysteresis}, anomalous Hall effect (AHE)~\cite{okamoto1983new,sato2011anomalous}, and magneto-optical properties~\cite{suran1999magnetic,berling2006accurate,cowburn1997new}.  The acquired data are then conventionally fitted using the torque balance method, which ultimately yields the anisotropy constants of the specimen.

In this study, an analysis of magnetic anisotropy and quasi-static magnetization dynamics in MRG thin-films, featuring a tetragonal crystal structure, is conducted through electrical transport measurement techniques (AHE).  MRG demonstrates a pronounced uniaxial out-of-plane anisotropy and a small yet significant four-fold in-plane anisotropy, which originates from substrate-induced compressive strain. MRG exhibits a substantial anomalous Hall effect, alongside a high magnetic anisotropy field and high Fermi-level spin polarization, a combination that enables direct probing of the anisotropy in MRG thin-films via electrical means. The manipulation of the magnetization vector ($\mathbf{M}$) of MRG within a 3D space, under the influence of a magnetic field, enables the examination of various anisotropy constants of the film. To characterize the equilibrium or dynamic response of the magnetization vector within an applied or induced effective field, accounting for the magnetic anisotropy of the sample is crucial. Generally, the equation of motion for magnetization is spatially non-uniform (described by a micromagnetic model) or, in a much simpler case, spatially uniform (explained by a macrospin model). This study employs the anomalous Hall effect to examine magnetic anisotropy in MRG, within the macrospin model framework, in combination with a distribution of hysterons, with finite magnetic viscosity and negligible interaction field.

This paper commences with a discussion of the sample preparation and characterization techniques employed in this study (section~\ref{sec:experimental}). Subsequently, section~\ref{sec:hysteresis_model} introduces the modeling of hysteresis in MRG under a classical Preisach model and the first-order reversal curves (FORC) method, wherein the validity of the macrospin model for MRG is established within the FORC and Preisach frameworks.  A comprehensive torque model for evaluating the anisotropy constants of MRG using AHE is explored in section~\ref{sec:torque_model}. Moreover, section~\ref{sec:combined_model} examines various intricate static and quasi-static magnetization dynamics of MRG through a 'combined' Preisach and torque model. Lastly, conclusions are drawn in section~\ref{sec:conclusion}.

\section{Experimental details}
\label{sec:experimental}
Epitaxial thin films of \mrg\, were fabricated using a DC magnetron sputtering system on a $ 10 \times 10 $ $\text{mm}^2 $ MgO (001) substrate. The films were co-sputtered in an inert environment (argon gas) from $\text{Mn}_{\text{2}}{\text{Ga}}$ and Ru targets onto the substrate, which was maintained at \SI{320}{\celsius}. Additional details regarding film growth and characterization can be found in a separate publication~\cite{Siewierska2021}. This study focuses on the $ x=0.9 $ stoichiometry and a film thickness of approximately $ \SI{40}{\nano\meter} $. The compensation temperature ($ \tcmp $) of this sample is considerably higher than room temperature, at $\tcmp \sim\SI{350}{\kelvin} $, as determined by \textsc{SQUID}$^{\circledR}$ magnetometry measurements. To prevent oxidation, the films were in-situ capped with approximately $ \sim \SI{3}{nm} $ of amorphous AlO$_x$, deposited at room temperature. The substrate-induced compressive strain ($c/a\approx1.02$) facilitated the out-of-plane magneto-crystalline anisotropy in the film. To investigate the transport properties, the films were patterned into micron-sized ($60\times 20$ $\mu m^2$) Hall bars, using UV photolithography and Ar-ion milling. A subsequent round of lithography and metal deposition was performed to establish the contact pads and to minimize the series resistance contribution of the corresponding contacts, consisting of Ti (\SI{5}{\nano\meter})/Au (\SI{50}{\nano\meter}).

The electronic transport properties were measured using the Quantum Design Physical Property Measurement System (\textsc{PPMS}$^{\circledR}$) in the temperature range of  $\SI{2}{\kelvin}\leq$ $T$ $\leq \SI{300}{\kelvin}$ and magnetic field strengths of $\left| {\mu_{0}H} \right| \leq \SI{14}{\tesla}$. The longitudinal and transverse voltages were measured by applying a lock-in demodulation technique at the first harmonic with low excitation frequency, typically $ f_{AC}\sim \SI{517}{\hertz}$, which was significantly smaller than the resonance frequencies of MRG. To determine the angular dependence of the resistivity, measurements were taken on a rotating platform within the PPMS, with an angular resolution of 0.01 deg. Additionally, First Order Reversal Curves (FORC) were measured using a field resolution of \SI{5}{\milli\tesla} at room temperature in a $\SI{1}{\tesla} $ \textsc{GMW}$^{\circledR}$ electromagnet.
\section{Results and Discussion}
\label{sec:result_and_discussion}
\subsection{ Hysteresis model}
\label{sec:hysteresis_model}
\begin{figure}[htb]
\centering
	\includegraphics[width= \linewidth]{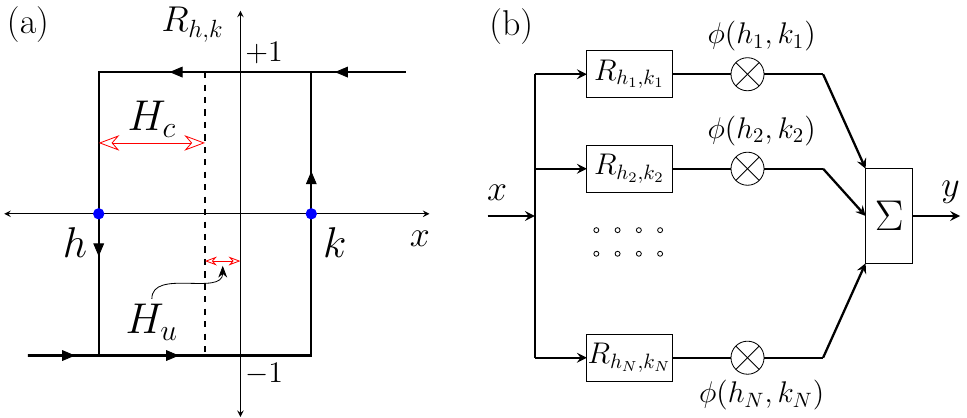}
    \subfloat{
    \label{fig:hysterons1}}
    \subfloat{
    \label{fig:hysterons2}}
    \caption{Illustration of the Preisach model. (a) A depiction of an elementary hysteron, a key component of the Preisach model, which exhibits unequal reversal fields with values $h$ and $k$. The hysteron's state is determined by the input variable $x$, as well as its history and strength. Hysteresis results from the collective interaction of numerous hysterons. (b) The discrete Preisach model of hysteresis, where a large number of hysterons are assumed to be connected in parallel, each with a corresponding weighting factor $ \phi(h,k) $. In this model, $x$ represents an arbitrary excitation variable (e.g., applied magnetic field), and $y$ signifies the resulting hysteretic physical property (e.g., magnetic moment).}
	\label{fig:hysteronsA}
\end{figure}

This section discussess the approach taken to model the switching of the magnetization (magnetic hysteresis) using the classical Preisach (hysterons) model. Hysteresis modeling has been an active area of research for decades, owing to both physical and mathematical interest. The magnetic hysteresis of ferromagnetic materials is the most famous example of hysteresis. It is widely accepted that the multiplicity of metastable states is the origin of hysteresis. Consequently, a micromagnetic model must be considered for hysteresis modeling.  In 1935, Preisach~\cite{preisach1935} proposed a classical micromagnetic mathematical approach to describe the hysteretic effect. The Preisach model (PM) employs a large number of interacting magnetic entities (referred to as hysterons),  each of which has a rectangular hysteresis loop (figure~\ref{fig:hysterons1}). These hysterons are characterized by the operator $ R_{h,k}(x) $, where $ x $ is an arbitrary input variable, such as an applied magnetic field. Hysteresis arises from the collective behavior of numerous hysterons, which switch fully at a discrete applied field. The value of $ R_{h,k}(x) $  relies on the applied field history. For instance, if the applied field $(x) $ starts from the saturation state ($x= \infty$), $ R_{h,k}(x) $ initiates at $ R_{h,k}(\infty)=1 $. The value of $ R_{h,k}(x) $ transitions to  $-1$ when the applied field falls below the value $h$, and $ R_{h,k}(x) $ returns to $+1$  when the field value exceeds $k$. Typically, the switching fields $h$ and $k$ are not identical.

The interaction field experienced by a hysteron is defined by $ H_u= (h+k)/2 $, resulting in an asymmetric elementary hysteron. In contrast, a hysteron with no interaction is symmetric. The coercive field of a hysteron is defined as $ H_c= (h-k)/2 $.  In a realistic sample, the hysteresis property is a weighted sum of a large number of hysterons, as described in equation~\ref{eq:hyst1}:
\begin{equation}\label{eq:hyst1}
y(x)  = \sum\limits_{i=1}^{N}\phi(h_i,k_i)R_{h_i,k_i}(x)\,.
\end{equation}
Here, the weighting factor $ \phi(h,k) $ represents the distribution of the switching fields $h$ and $k$ and is commonly referred to as the switching field distribution (SFD) or hysteron distribution (HD). Figure~\ref{fig:hysterons2} illustrates a schematic representation of the Preisach model.

In the continuum limit, the discrete model is transformed into the following expression:
\begin{equation}\label{eq:hyst2}
y(x)=\iint\limits_{k\geq h}\phi(h,k)R_{h,k}(x) \,\mathrm{d}h\,\mathrm{d}k\, ,
\end{equation}
where, $ x $ is an arbitrary variable (e.g., applied magnetic field) and $ y $ is the resultant  hysteresis output (e.g., magnetic moment, anomalous Hall voltage, etc.). 
The most challenging aspect of the Preisach model involves uniquely defining the distribution function $ \phi(h,k) $. Nevertheless, for an assembly of weakly interacting hysterons, it is possible to assume that the distribution function $ \phi(h,k) $ follows a specific statistical distribution. Common choices include the Gaussian function~\cite{DellaTorre1986,Kadar1989}, Gauss-Lorentzian function~\cite{Fuzi2003}, and Lognormal-Gaussian distribution function~\cite{Henze2002}, among others. However, this approach faces the issue of lacking justification for selecting one particular distribution over others~\cite{Henze2002}. An alternative method entails using a linear combination of a set of functions as a basis. The drawback of this approach is the requirement of a large set of basis functions and their coefficients to obtain a Preisach distribution with a relatively continuous output (y)~\cite{Galinaitis2001}, which rapidly strains computational capabilities, even for modern computers.

The first-order reversal curves (FORC) method offers an experimental technique for obtaining a unique Preisach distribution, as long as the sample of interest meets the necessary and sufficient Mayergoyz conditions~\cite{mayergoyz1986mathematical}. The FORC method is both easily achievable experimentally and highly reproducible, given that it begins by saturating the sample each time. It has been employed to examine various magnetic systems, such as permanent magnets~\cite{chiriac2007experimental,chen2014soft}, geological samples~\cite{roberts2000first,muxworthy2007first}, nanowires~\cite{beron2006first,beron2008magnetic}, and more. Moreover, FORC can differentiate between interacting and noninteracting single domain (SD), pseudo single-domain (PSD), and multi-domain (MD) systems~\cite{pike1999characterizing,roberts2000first}. In fact, the FORC method can be extended to any system exhibiting hysteresis behavior, including ferroelectric samples~\cite{stancu2003first,ramirez2009first}. Additionally, FORC studies on certain magnetic systems can be complemented by AHE measurements, where electrical probing presents a decisive advantage over standard magnetic moment measurements~\cite{diao2012magnetization}.

\begin{figure*}[htb]
\centering
    \begin{minipage}[c]{0.40\linewidth}
    \centering
    \subfloat[\label{fig:FORC_loop}]{
    \includegraphics[width=\linewidth]{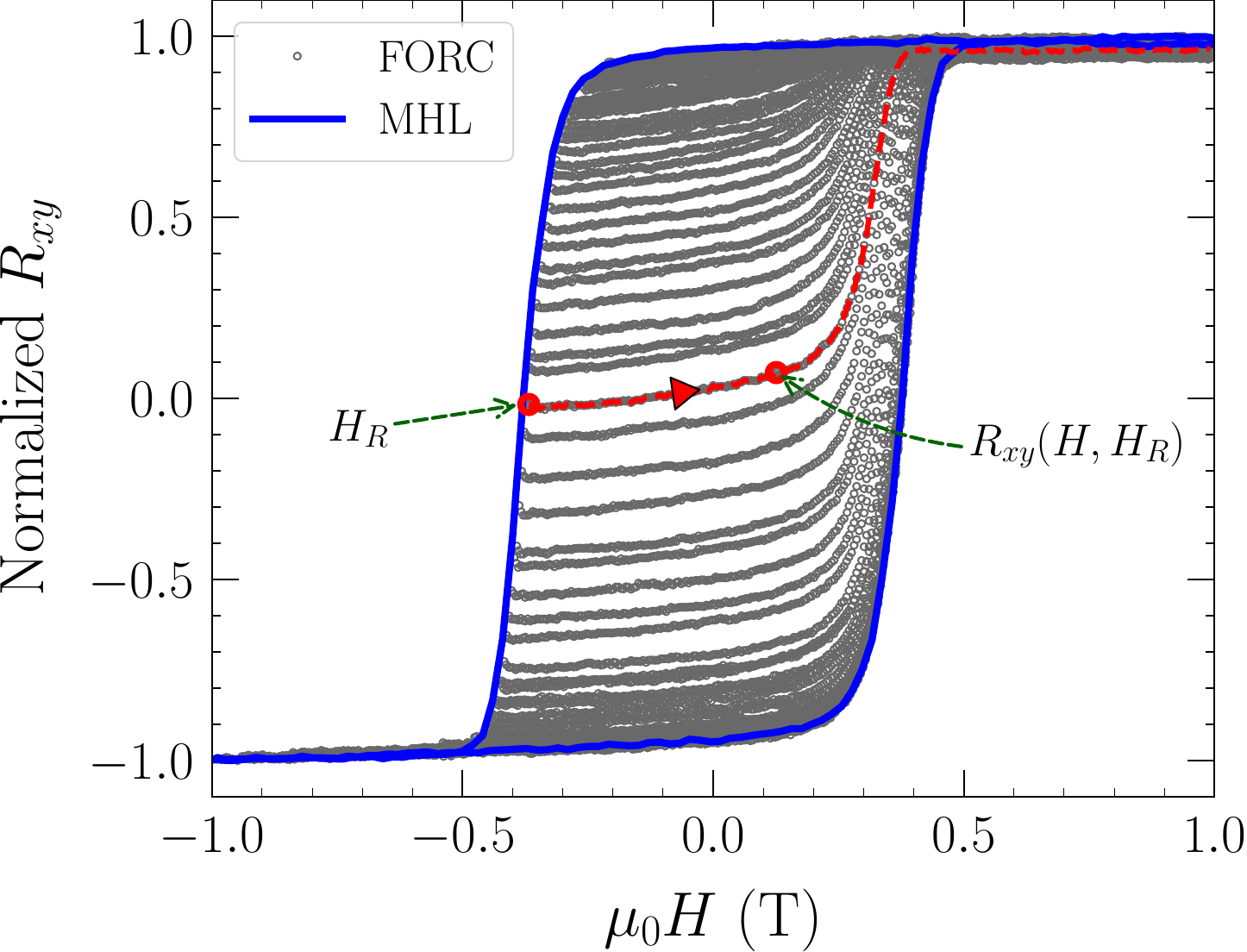}
    }
    \end{minipage}
    \hfil
    \begin{minipage}[c]{0.40\linewidth}
    \centering
    \subfloat[\label{fig:FORC}]{
    \includegraphics[width=\linewidth]{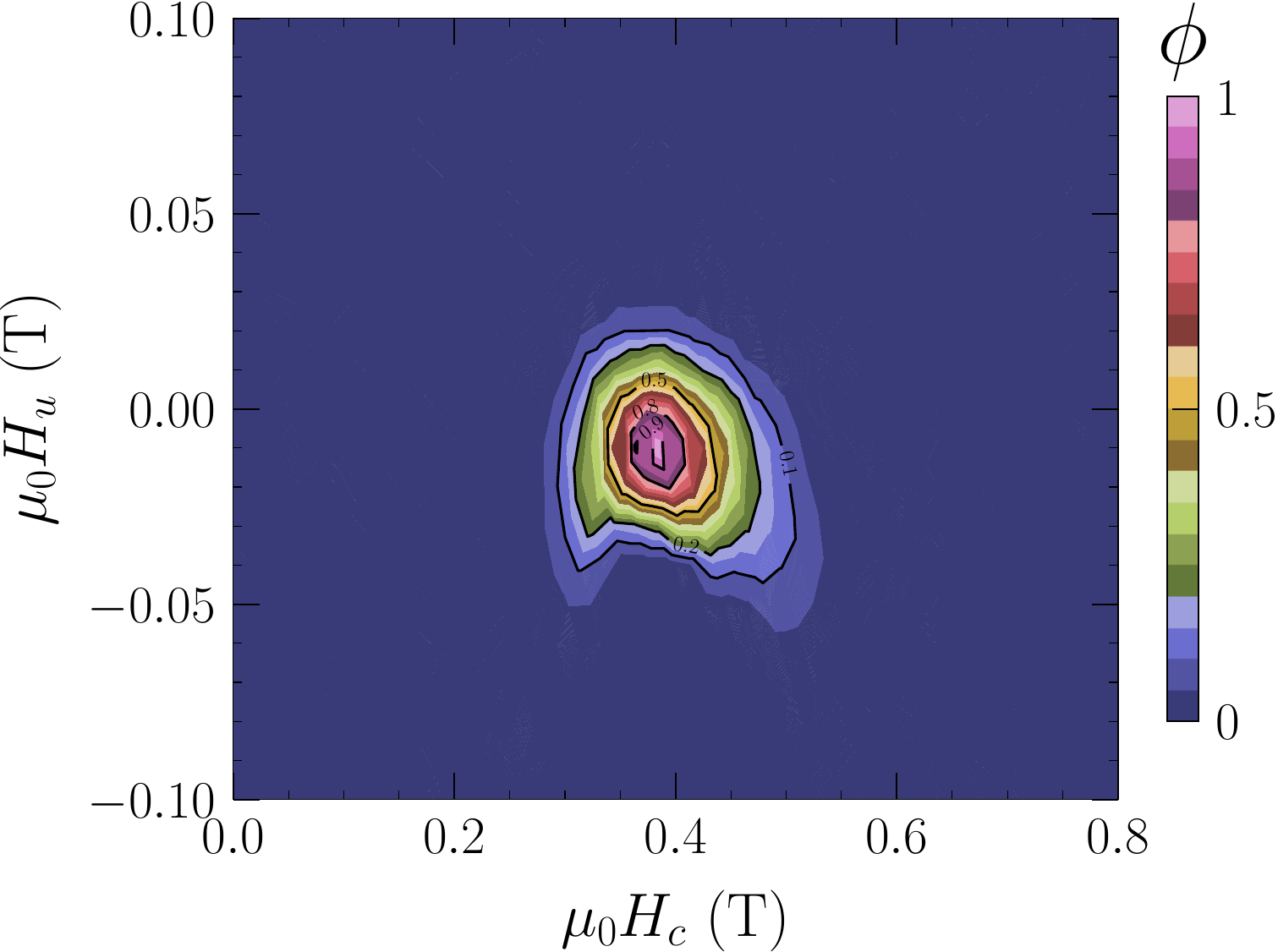}
    }
    \end{minipage}
    \hfil
    \begin{minipage}[c]{0.40\linewidth}
    \centering
    \subfloat[\label{fig:FORC_Hu}]{
    \includegraphics[width=\linewidth]{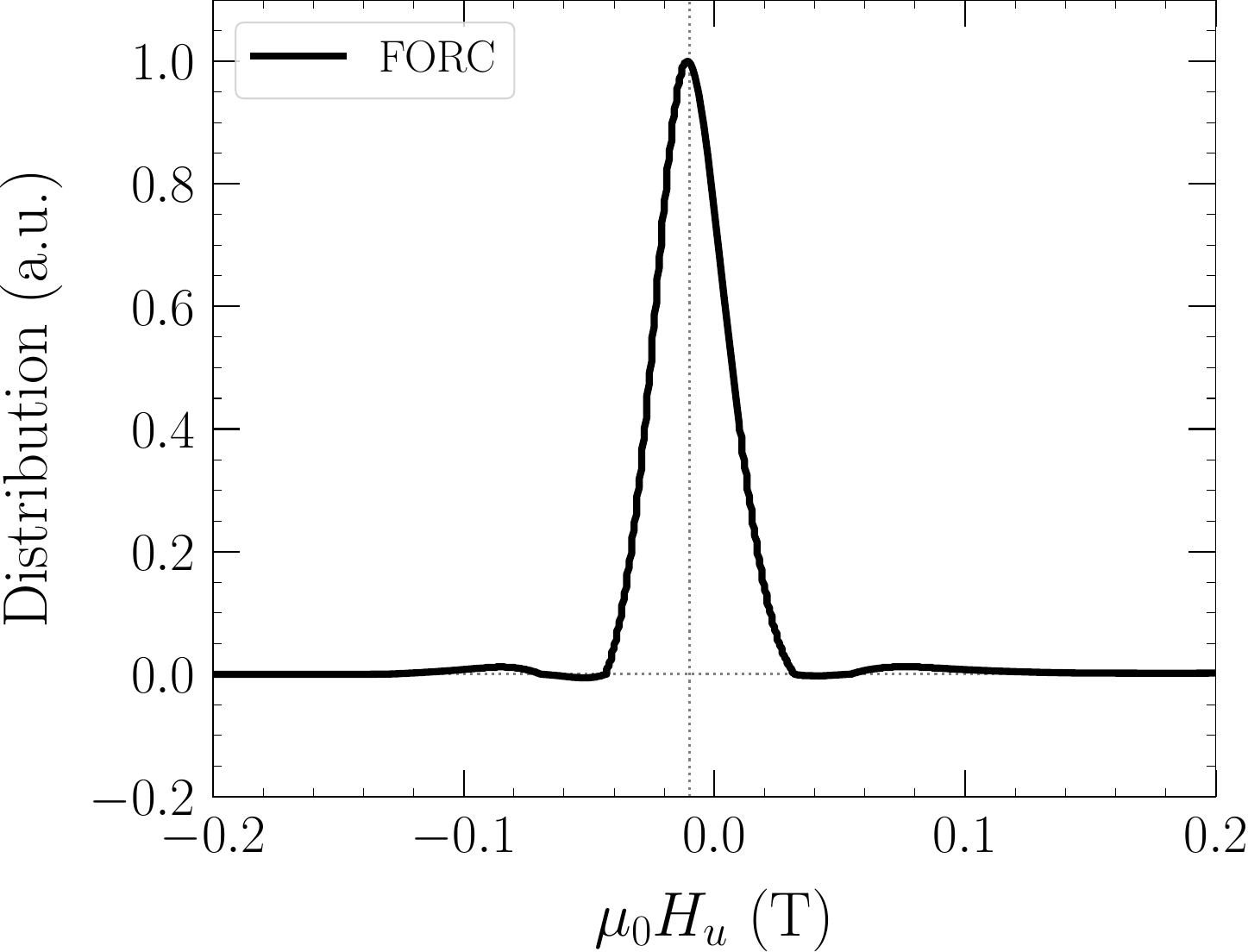}
    }
    \end{minipage}
    \hfil
    \begin{minipage}[c]{0.40\linewidth}
    \centering
    \subfloat[\label{fig:FORC_Hc}]{
    \includegraphics[width=\linewidth]{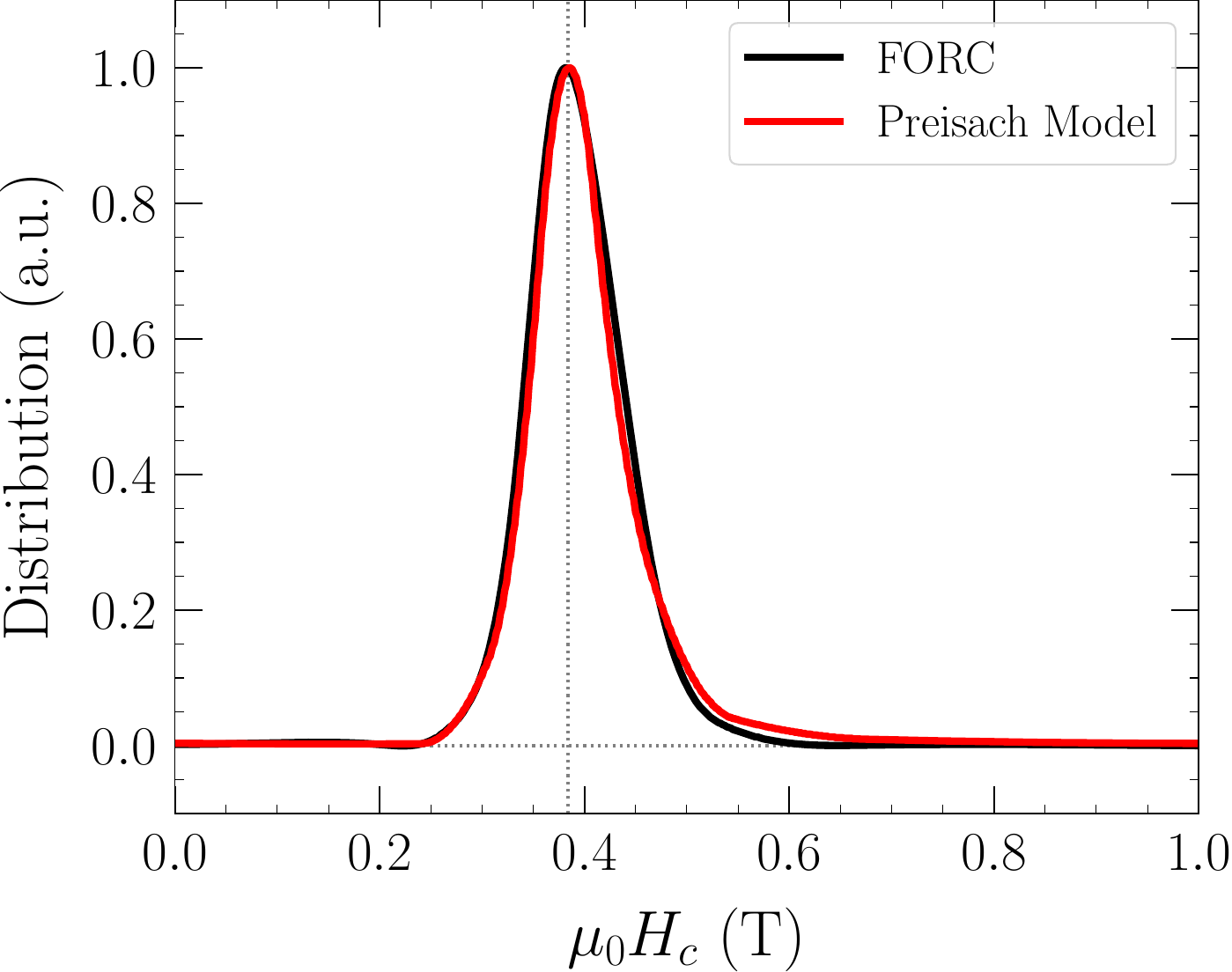}
    }
    \end{minipage}
    \caption{
    Investigation of FORC and Preisach model for MRG. (a) FORC measurements of MRG using the AHE. The main hysteresis loop (MHL) is depicted as a blue solid curve, acquired when the magnetic field is swept from \protect\SI{\pm1}{\tesla}. A minor AHE curve is subsequently obtained by starting from the saturation point and returning the magnetic field value to a lower field on the MHL, referred to as the reversal field ($H_R$). The field is then brought back to the saturation point, forming a single FORC curve (dashed red curve). This process is repeated for numerous uniformly spaced values of $H_R$ and applied magnetic field ($H$), resulting in the FORC diagram covering the area within the MHL (black data points). (b) FORC distribution for MRG, derived using equation~\ref{eq:FORC_AHE}. The distribution is attained by fitting the FORC grid utilizing a local second-order polynomial. The distribution is presented in transformed interaction field ($H_u$) and coercivity field ($H_c$) axes for convenience. (c) Interaction field ($H_u$) distribution of MRG derived from the FORC distribution. A narrow $H_u$ distribution, centered at $\mu_{0}H_u = -0.01~\text{T}$, highlights the absence of long range interactions (dipolar) between the hysterons comprising the MRG. (d) Resultant coercive field distribution $H_c$, centered at $\mu_{0}H_c = 0.39~\text{T}$, from the FORC distribution (black line) and the estimated coercive field distribution as per the Preisach model (red line). The congruence between the two curves validates the proposed Preisach model for the MRG system.
    }
	\label{fig:FORC_A}
\end{figure*}

The FORC measurement using AHE commence by saturating the sample in a sufficiently high positive magnetic field. Subsequently, the field is decreased to a lower field value on the main hysteresis loop (MHL), referred to as the reversal field ($ H_R $), and the Hall resistance $ R_{xy}(H,H_R) $ is measured by sweeping the applied field $H$ back to the saturation field. The resulting AHE resistance, $ R_{xy}(H,H_R) $, constitutes a minor curve within the MHL (figure~\ref{fig:FORC_loop}). This procedure is repeated for numerous uniformly spaced values of $ H_R $ and $ H $.
The FORC distribution is acquired through the second-order mixed derivative, as defined by equation~\ref{eq:FORC_AHE}\,:

\begin{equation}\label{eq:FORC_AHE}
	\phi(H,H_R)= -\frac{1}{2}\frac{\partial^2}{\partial H \partial H_R}[R_{xy}(H, H_R)]\, .
\end{equation}

The FORC distribution was assessed through the application of a locally fitted second-order polynomial surface. A gradient smoothing factor was incorporated into the algorithm to suppress numerical artifacts. Conventionally, FORC diagrams are depicted in terms of the coercivity field ($H_c$) and the interaction field ($H_u$), which can be derived using  $ H_c= (H_R-H)/2 $  and $ H_c= (H_R+H)/2 $. The resulting FORC diagram is displayed in figure~\ref{fig:FORC}, where a central ridge is observed around $\mu_0H_u =  -0.01~\text{T}$ and $\mu_0H_c= 0.39~\text{T}$. Figure~\ref{fig:FORC_Hu} illustrates the local interaction field distribution of the MRG. A narrow distribution of $H_u$, with a central point at $\mu_{0}H_u = -0.01~\text{T}$, emphasizes the lack of any significant interactions (dipolar, etc.) between the elementary units (hysterons) that comprise the MRG. Consequently, in the absence of inter-particle interactions, the overall system can be reasonably approximated using the Stoner-Wohlfarth (SW) model~\cite{stoner1948mechanism}.

The coercive field distribution of the FORC diagram is depicted in figure~\ref{fig:FORC_Hc}, with the peak of the distribution centered at $\mu_0H_c= 0.39~\text{T}$. In the absence of interactions, coercive field distribution also represents the switching field distribution (SFD) of hysterons. A statistical analysis of SFD was conducted within the framework of the Preisach model. For this analysis, a pseudo-Voigt distribution is employed, defined as:

\begin{equation}\label{eq:Voigt}
	V(H_c,H_{c_0},\Gamma)= \eta G(H_c,H_{c_0},\Gamma) + (1-\eta) L(H_c,H_{c_0},\Gamma)\, ,
\end{equation}
where, $G(H_c,H_{c_0},\Gamma)$ and $L(H_c,H_{c_0},\Gamma)$ are normalized Gaussian and Lorentzian function. $\Gamma$ is the common FWHM and $H_{c_0}$ is peak center. $\eta$ ($0\leq\eta \leq1 $) serves as a weighting factor that transitions the overall profile between pure Gaussian and pure Lorentzian distributions by adjusting the factor from 1 to 0, respectively.

The coercive field distribution within the FORC diagram can be suitably fitted using equation~\ref{eq:Voigt}. The elongated tail of the coercivity distribution is attributable to the magnetic viscosity resulting from the thermal fluctuations of metastable states. In MRG, magnetic viscosity predominantly stems from the rotation of the magnetization vector, as contributions from domain wall motion are substantially hindered by defects and disorder present within the film~\cite{teichert2021magnetic}.  Therefore, viscosity can be expressed as the sum of exponentially decaying metastable states. Convolution of these states with the pseudo-Voigt function results in the Preisach distribution or switching field distribution (SFD), as demonstrated in equation~\ref{eq:hys_mod}:
\begin{equation}\label{eq:hys_mod}
\resizebox{\linewidth}{!}{%
	$\displaystyle D(H_c,H_{c_0},\Gamma , \tau)=\int\limits_{-\infty}^{\infty}V\left[ (H_c-\xi),H_{c_0},\Gamma\right] \frac{1}{\tau}\left[ \exp\left(\frac{-\xi}{\tau}\right)\right]\mathrm{d}\xi\, ,
$}
\end{equation}  
here, $\tau$ represents the magnetic viscosity parameter, measured in units of magnetic field.
Figure~\ref{fig:FORC_Hc} provides clear evidence of a strong agreement between the experimentally obtained coercive field distribution under FORC method and the theoretical prediction provided by the Preisach distribution (equation~\ref{eq:hys_mod}). This finding supports the conclusion that the distribution described by equation~\ref{eq:hys_mod} can be safely considered as a unique Preisach distribution of the MRG samples. It is worth noting that appropriate normalization methods (amplitude or arial) must be implemented in order to accurately signify the deterministic switching of hysterons. These findings not only contribute to a better understanding of the switching behavior of MRG samples, but also have important implications for the development of more robust models for other similar systems.

\begin{figure}[htbp]
	\centering
	\begin{minipage}[c]{0.80\linewidth}
		\centering
		\subfloat[\label{fig:P_300K}]{%
			\includegraphics[width=\linewidth]{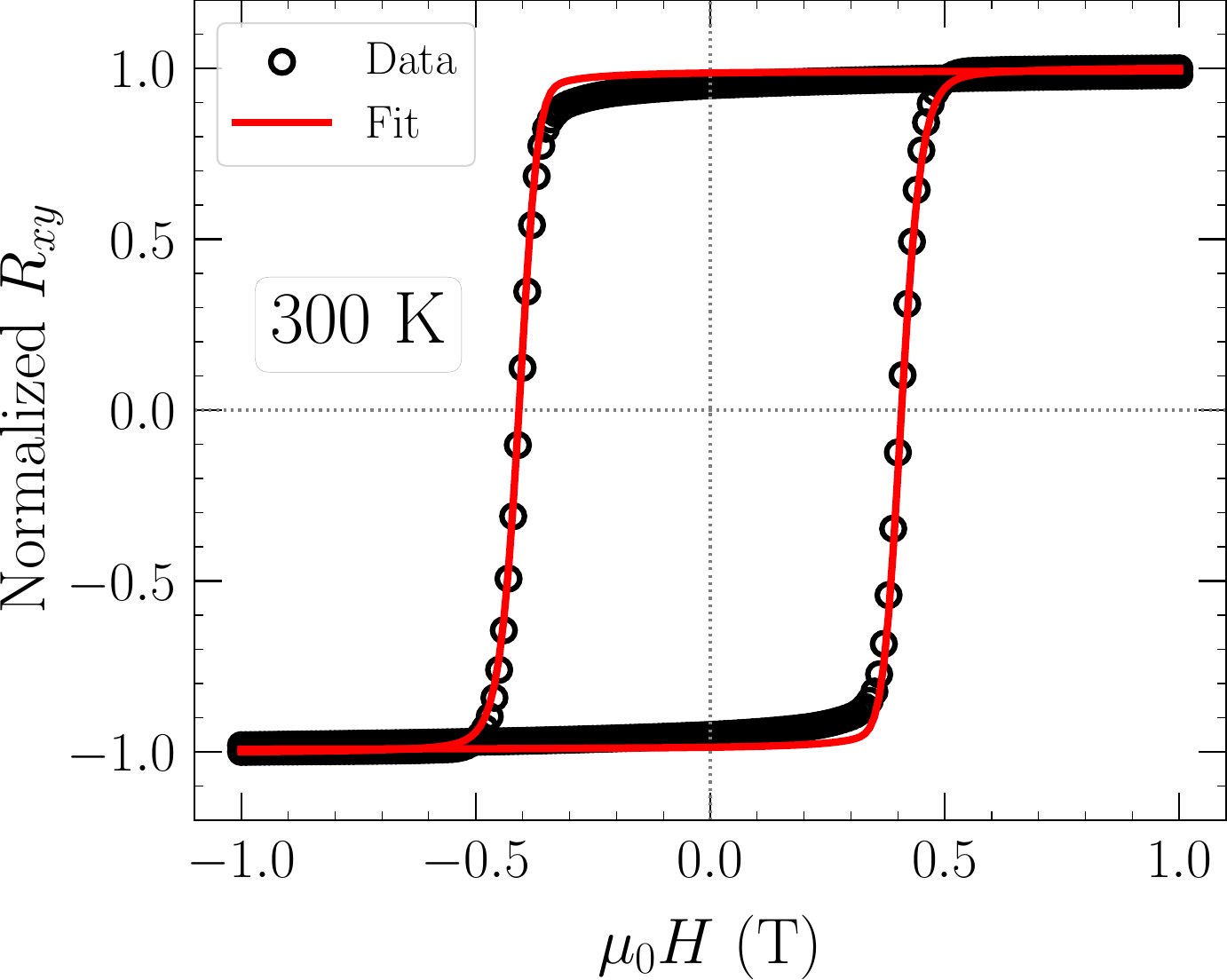}}
		\vspace{-0.1cm}
	\end{minipage}
	\begin{minipage}[c]{0.79\linewidth}
		\centering
		\hspace{0.1cm}
		\subfloat[\label{fig:Hy_300K}]{%
			\includegraphics[width=\linewidth]{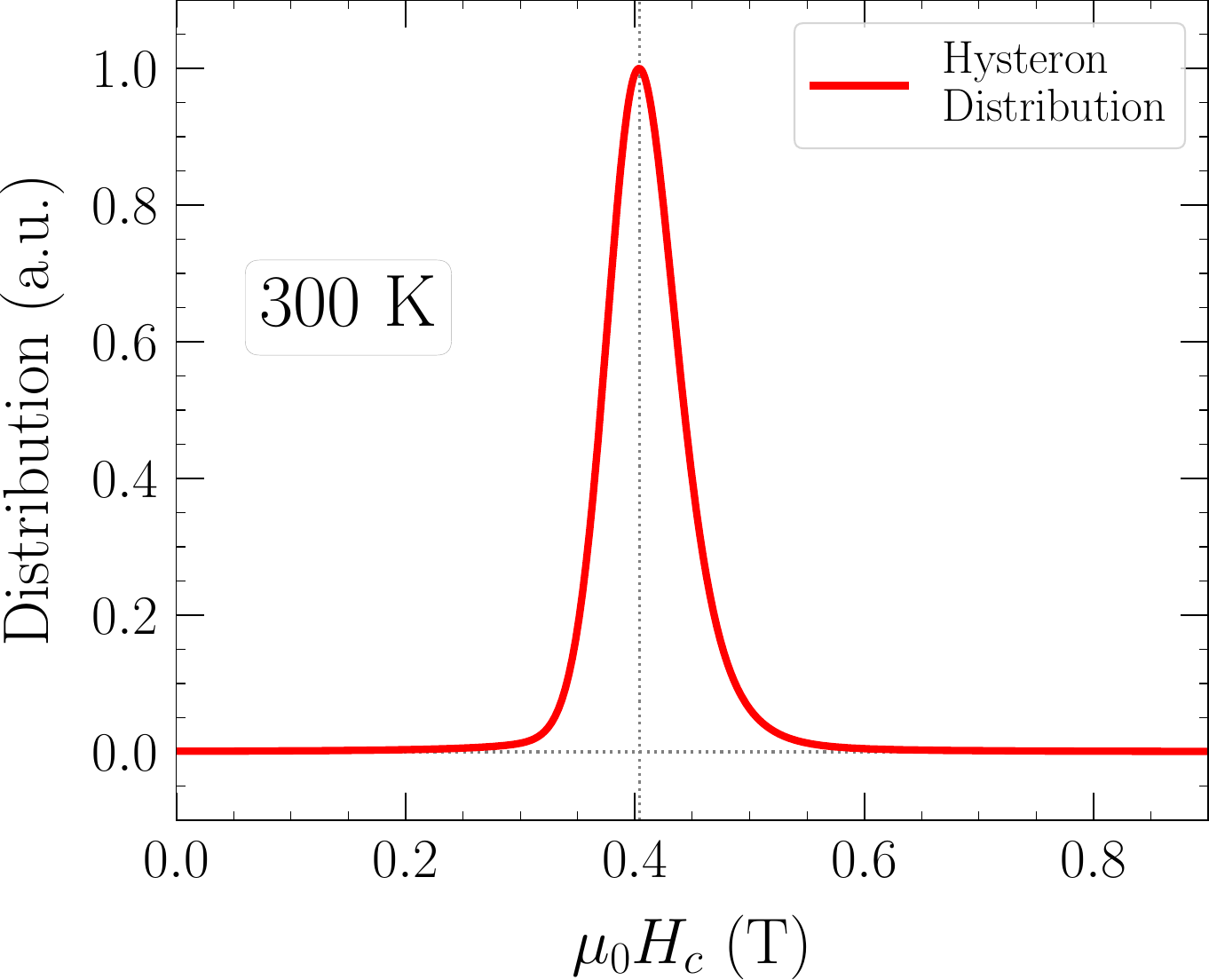}}
	\end{minipage}
    \caption{\label{fig:hysteresis_fit_1}Hysteresis of MRG characterized using the Preisach model. (a) The experimentally acquired AHE hysteresis loop (black circles) recorded at \SI{300}{\kelvin}, obtained when the applied magnetic field was swept perpendicular to the film plane. The estimated hysteresis (red line), derived from the Preisach model (equations~\ref{eq:hyst2} and~\ref{eq:hys_mod}), demonstrates a high degree of agreement with the experimental data, effectively capturing the key features of the observed hysteresis loop. (b) The Preisach distribution (PD) for this MRG sample, which depicts the distribution of elementary hysterons, is centered at \SI{0.40}{\tesla} exhibits a notably narrow distribution.}
\end{figure}

Upon obtaining the requisite hysteron distribution, a hysteresis curve for the MRG can be seamlessly derived by integrating this distribution into the Preisach model (equation~\ref{eq:hyst2}.). Figure~\ref{fig:P_300K}  demonstrates a remarkable congruence between the experimental AHE hysteresis data obtained at \SI{300}{\kelvin} and the fit generated through the Preisach model, with the corresponding resultant Preisach distribution presented in figure~\ref{fig:Hy_300K}. Furthermore, this model has been expanded to encompass out-of-plane hysteresis measurements of MRG at various other temperatures.

\begin{figure}[htb]
\centering
 \includegraphics[width=\linewidth]{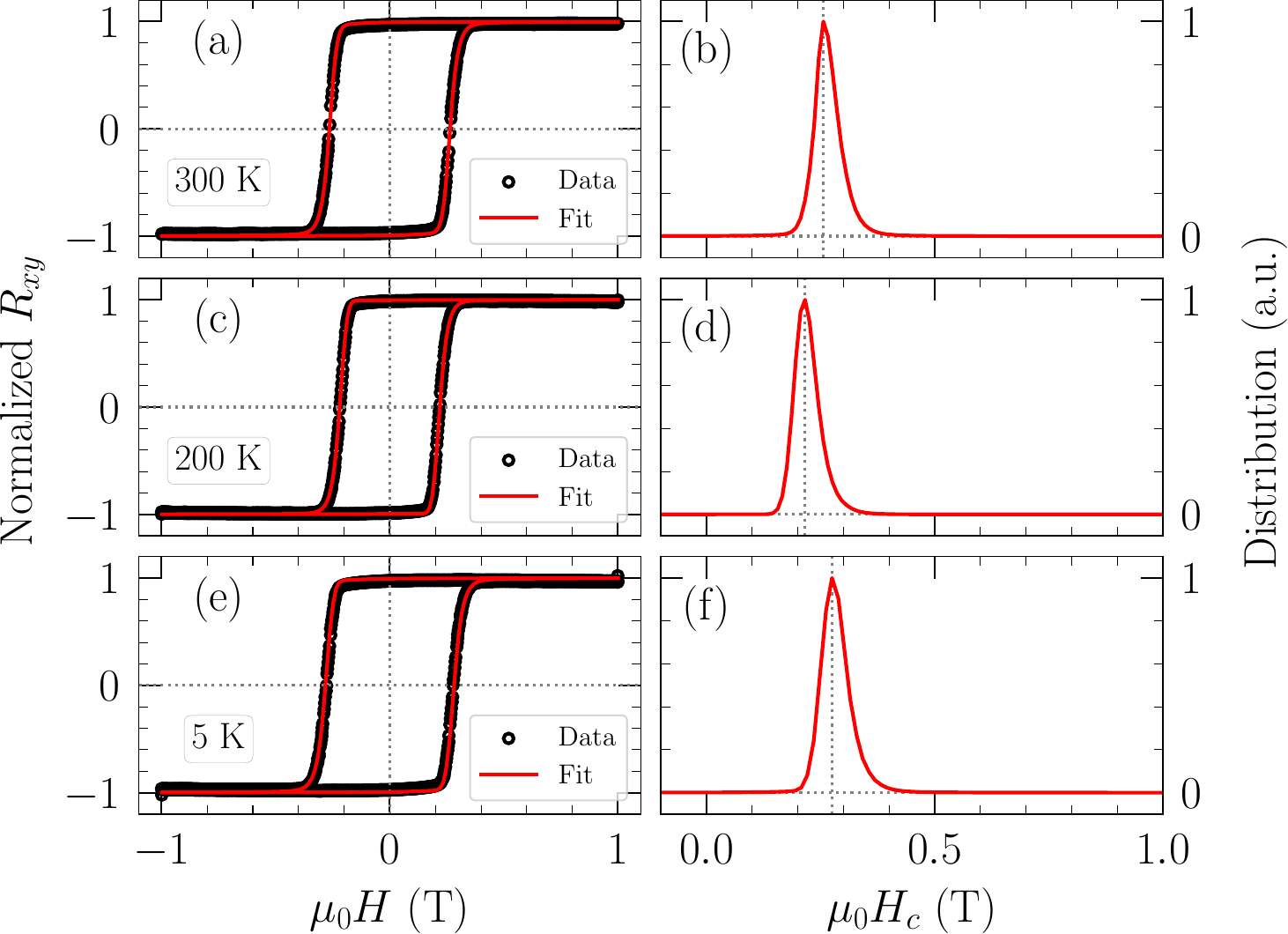}
\caption{Temperature-dependent hysteresis of MRG characterized using the Preisach model. The experimentally recorded AHE hysteresis loops (black circles) measured at \SI{200}{\kelvin}, \SI{100}{\kelvin}, and \SI{5}{\kelvin} are presented in panels (a), (c), and (e) respectively, with their corresponding Preisach model estimations (red lines) capturing the experimental data in great detail. The estimated Preisach distribution of each model are depicted in panels (b), (d), and (f), providing insights into the hysteron distributions and their temperature-dependent behavior.}
\label{fig:hysteresis_fit}
\end{figure}

\begin{figure}[htb]
\centering
    \begin{minipage}[b]{0.80\linewidth}
    \centering
    \subfloat[\label{fig:H_peak}]{
    \includegraphics[width=\linewidth]{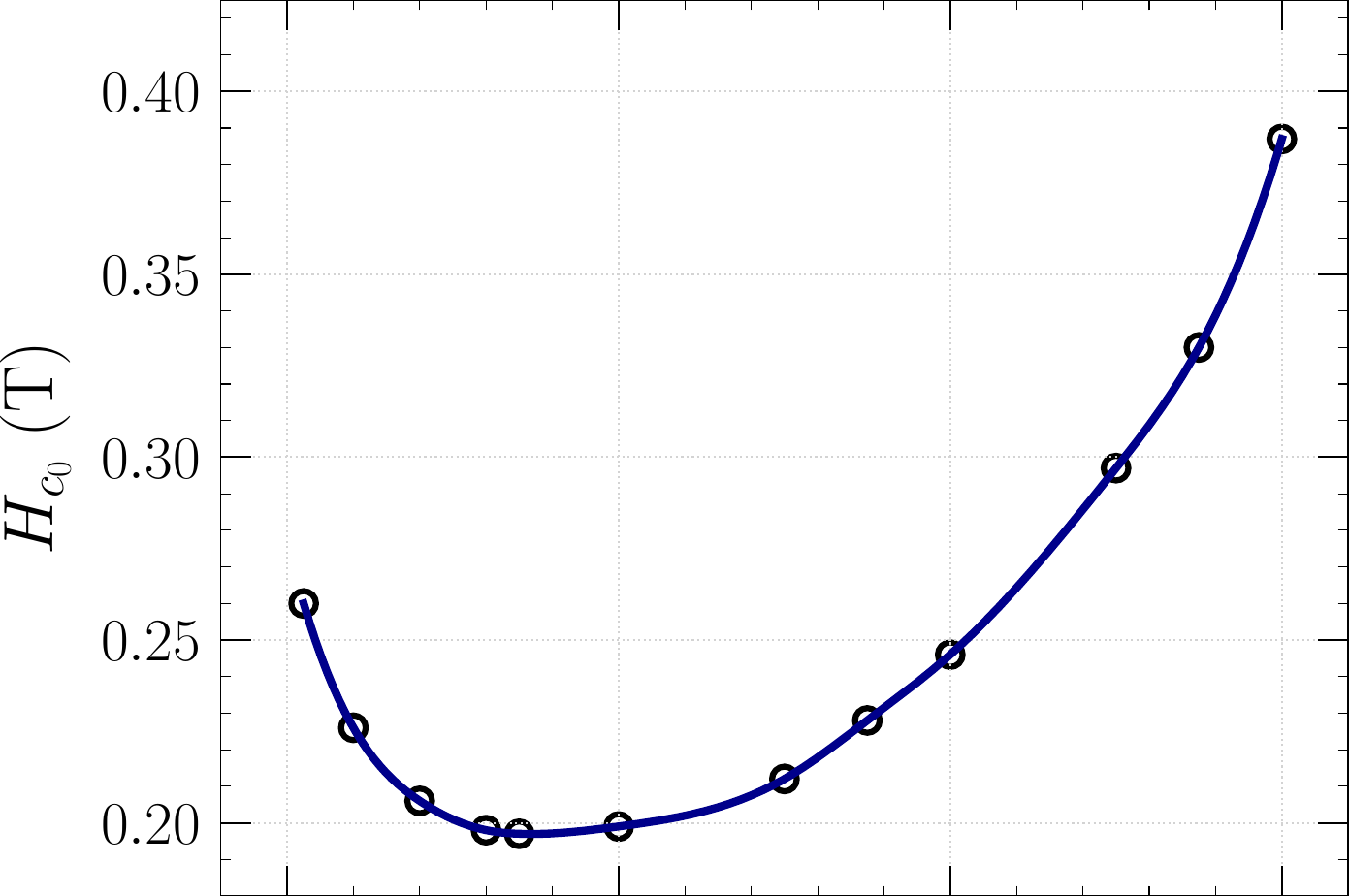}
    }
    \end{minipage}
    \begin{minipage}[b]{0.80\linewidth}
    \centering
    \subfloat[\label{fig:viscosity}]{
    \includegraphics[width=\linewidth]{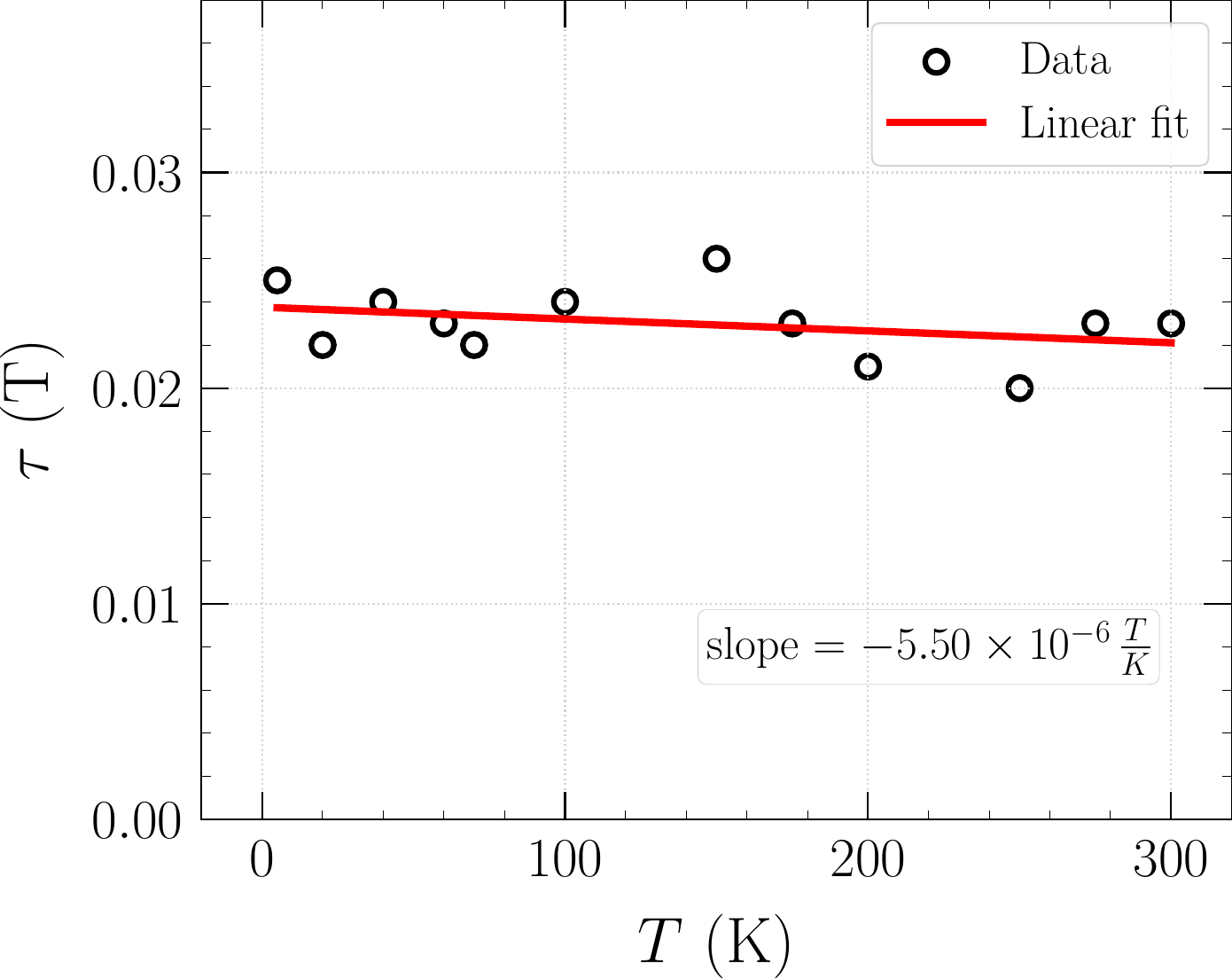}
    }
    \end{minipage}
	\caption{Temperature-dependent analysis of coercivity and magnetic viscosity in MRG. (a) The center-point ($H_{c_0}$) of the Preisach distribution, which also represents the sample's coercivity, as a function of temperature. The value of $H_{c_0}$ increases for both high and low temperature ranges. The increase at higher temperatures is attributed to the approach towards the compensation temperature ($T_{comp} = \SI{375}{\kelvin}$), while at lower temperatures, the rise in the anisotropy constant leads to an increase in the central-point value. The solid line serves as a guide for the data points. (b) Viscosity parameter ($\tau$) as a function of temperature. This relationship suggests that magnetic viscosity remains approximately independent within the measured temperature range, indicating that the magnetic domains for MRG are essentially frozen over a wide temperature range.
    }
	\label{fig:hysteresis_par}
\end{figure}

Figure~\ref{fig:hysteresis_fit} illustrates the AHE hysteresis loops and corresponding Preisach distributions at select temperature values, such as \SI{200}{\kelvin}, \SI{100}{\kelvin}, and \SI{5}{\kelvin}. As evidenced by these results, the model captures the experimental intricacies with remarkable precision, thereby underscoring its ability to accurately represent the extensive range of hysteresis observed in MRG using with only a limited number of parameters ($H_{c_0},\Gamma~ \text{and}\,\tau$). Additional insights into the magnetic properties can also be gleaned from this model.
Figure~\ref{fig:hysteresis_par} depicts the variations in the center point ($H_{c_0}$) and magnetic viscosity ($\tau$) as functions of temperature for hysteresis curves measured at diverse temperature range. A notably weak dependency of the magnetic viscosity parameter on temperature is observed, which can be ascribed to the dominant influence of anisotropy on the overall energy landscape of MRG. Consequently, any weaker thermodynamic fluctuations exert negligible impact on the static dynamics of the magnetization, causing the magnetic domains of MRG to remain frozen over a wide temperature range.

The relationship between the center point ($H_{c_0}$), which is also known as the sample's coercivity, and temperature is characterized by two distinct regimes. At elevated temperatures, the coercive field experiences an increase due to the diminishing net moment as it approaches the compensation point ($T_{comp} = \SI{375}{\kelvin}$); conversely, at lower temperatures, the rise in effective anisotropy prevails.

\subsection{Torque model}
\label{sec:torque_model}
In the investigation of magnetization dynamics, employing the macrospin approximation serves as a highly effective approach for analysis. In this approximation, the spatial variation of the magnetization remains constant throughout the equation of motion. The static and quasi-static magnetization dynamics of MRG can be accurately represented under the macrospin approximation, as it accounts for the absence of hysteron interaction, which is clearly illustrated in figure~\ref{fig:FORC_Hu}. The torque model is constructed based on the macrospin approximation, where the equilibrium direction of the magnetization is determined by counterbalancing the torque that arises from anisotropy fields with the Zeeman torque. For the tetragonal MRG system, the torque balance equation can be efficiently derived from the magnetic anisotropy free energy expression, in which $ \theta_M $ and $ \varphi_M $ represent the polar and azimuthal angles of the magnetization vector $ \mathbf{M} $:
\begin{eqnarray}\nonumber\label{eq:anisoeng}
E=K_1\sin^2(\theta_M)+K_2\sin^4(\theta_M)&+&K_3\sin^4(\theta_M)\cos(4\varphi_M)\\&&-\mu_0\textbf{H}\cdot\textbf{M}\, .
\end{eqnarray}
Here, the first and second-order uniaxial out-of-plane anisotropy constants are denoted by $ K_1 $ and $ K_2 $, respectively, while $ K_3 $ signifies the four-fold in-plane anisotropy constant. By evaluating the extrema of equation~\ref{eq:anisoeng} with respect to $ \theta_M $ and $ \varphi_M $, the equilibrium magnetization direction can be determined. For instance, the polar equilibrium position can be ascertained by solving the subsequent equation:
\begin{eqnarray}\nonumber\label{eq:anisoeng_2}
\frac{\partial E}{\partial \theta_M}&=&2K_1+\left[ 4K_2+4K_3\cos(4\varphi_M)\right]\sin^2(\theta_M)\\ &&-\dfrac{\mu_0HM\sin(\theta_H-\theta_M)}{\sin(\theta_M)\cos(\theta_H)}=0\, .
\end{eqnarray}  
In the aforementioned equation, it is assumed that the in-plane anisotropy ($ K_3 $) is relatively weak; therefore, $ \textbf{M} $  adheres to the applied magnetic field ($ \textbf{H} $) along the azimuthal direction with a slight delay, i.e., $ \varphi_M  \approx \varphi_H$. Here, the polar angle  and the azimuthal  angle of the applied magnetic field are represented by $ \theta_H $ and $\varphi_H$, respectively. In this work, we utilize the anomalous Hall effect (AHE) to examine the anisotropy constants, which is particularly sensitive to the out-of-plane component of the Mn$ ^{4c} $  moment, therefore:

\begin{equation}
\begin{split}
V_{xy}\propto {\textbf{M}}\cos(\theta_M)\, ,\\
\Longrightarrow \cos(\theta_M)=\dfrac{V_{xy}}{V_{xy}^N}=v_z\, ,
\end{split}
\end{equation}
where, $ V_{xy}^N $ denotes the AHE voltage when the magnetization ($ \textbf{M} $) is aligned with the normal to the sample ($ \theta_M  = 0$), and $ v_z $ represents the normalized AHE voltage. Consequently, the equilibrium condition (equation~\ref{eq:anisoeng_2}) is reduced to:
\begin{eqnarray}\nonumber\label{eq:GST}
\frac{2K_1}{M}+\left( \frac{4K_2}{M}+\frac{4K_3}{M}\cos(4\varphi_M)\right)\left( 1-v_z^2\right)\\
&&\hspace{-35mm}=\left( \frac{\mu_0H\sin(\theta_H-\theta_M)}{v_z\sqrt{1-v_z^2}}\right)\,.
\end{eqnarray}
To determine the values of $ K_1 $, $ K_2 $, and $ K_3 $, rotational scans were conducted in various geometric configurations.

\begin{figure}[htb]
\centering
	\hspace{-1cm} 
    \begin{minipage}[c]{0.795\linewidth}
    \centering
    \subfloat[\label{fig:OHE}]{
    \includegraphics[width=\linewidth]{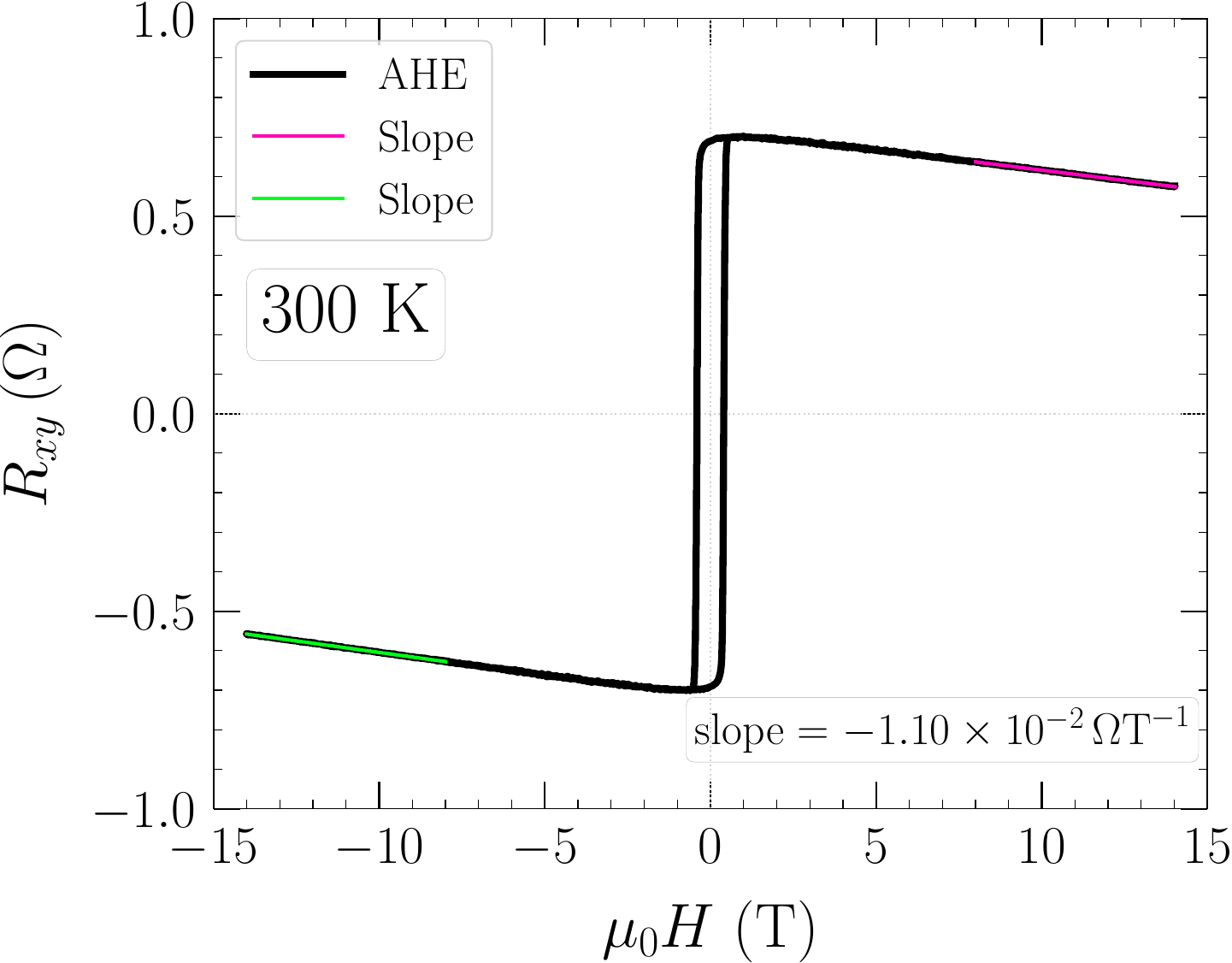}
    }
    \vspace{-0.1cm}
    \end{minipage}
    \begin{minipage}[c]{0.95\linewidth}
    \centering
    \subfloat[\label{fig:planner}]{
    \includegraphics[width=\linewidth]{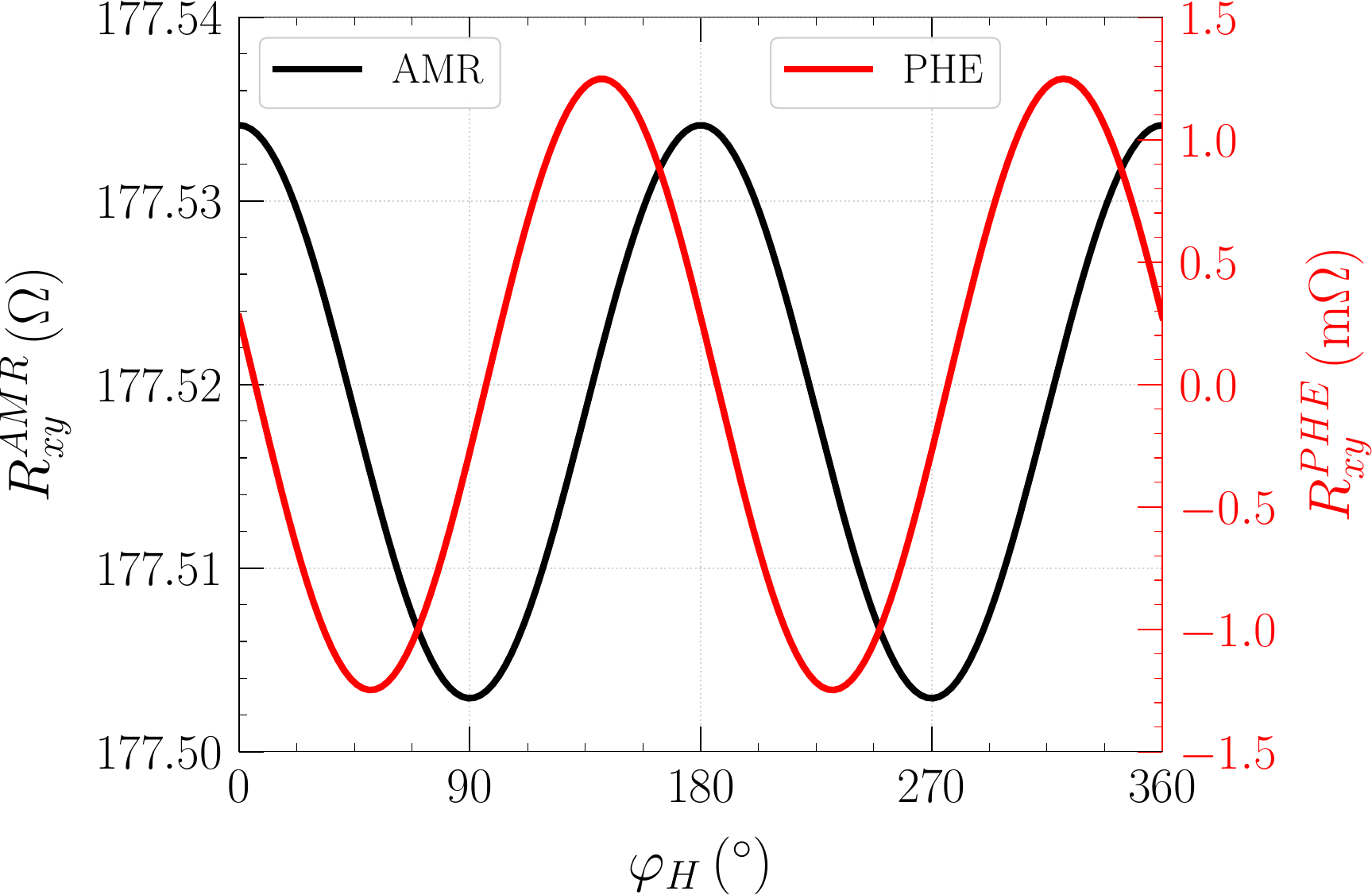}
    }
    \end{minipage}
	\caption{AHE, AMR, and PHE measurements of MRG at \SI{300}{\kelvin} using a micron-sized  ($60\times 20$ $\mu m^2$) Hall bar. (a) AHE is recorded with the magnetic field applied perpendicular to the sample within a range of \SI{\pm14}{\tesla}. The AHE hysteresis loop exhibits a coercivity value of \SI{0.40}{\tesla}. The ordinary Hall effect (OHE) contribution of MRG is evaluated by calculating the slope of the curve at high magnetic fields ($|\mu_0H| > \SI{8}{\tesla}$). (b) AMR and PHE of MRG are measured at \SI{300}{\kelvin} when a \SI{1.9}{\tesla} magnetic field is rotated within the sample plane. The PHE (of the order of a few milliOhm) has a contribution three orders of magnitude smaller than that of AHE. Consequently, the PHE contribution from the transverse Hall effect can be safely disregarded.
    }
	\label{fig:AMR_PHE}
\end{figure}

It is important to recognize that the recorded transverse resistance consists of five distinct contributions, which include: the ordinary Hall effect (OHE), the anomalous Hall effect (AHE), the planar Hall effect (PHE), the ordinary Nernst effect (ONE), and the anomalous Nernst effect (ANE). These contributions are represented in equation~\ref{eq:halleffect}~\cite{RevModPhys.82.1539,PhysRevLett.99.086602}:
\begin{eqnarray}\label{eq:halleffect}
	&&\hspace{-10mm}R_{xy}= R_{xy}^{OHE}+R_{xy}^{AHE}+R_{xy}^{PHE}+R_{xy}^{ONE}+R_{xy}^{ANE}\, .
\end{eqnarray}
In the context of MRG, the ordinary Nernst effect ( $R_{xy}^{ONE} $ ) and the anomalous Nernst effect ($R_{xy}^{ANE}$ ) were effectively minimized by employing an exceedingly small input bias current signal ($ I_{RMS}\approx\SI{50}{\micro\ampere}$). This approach ensured the absence of any significant thermal gradient within the observed sample. To further mitigate the temperature gradient across the Hall bar, temperatures were stabilized using a helium partial pressure ($ P\sim \SI{100}\,\text{Torr}$) within a PPMS tool. The sample was carefully rotated at a very slow rate to minimize temperature destabilization that could arise from friction within the sample's rotator gears.
The ordinary Hall effect ($R_{xy}^{ONE} $ ) was determined by measuring the slope of the AHE at high magnetic fields ($|\mu_{0}H|>8~\text{T}$), as illustrated in figure~\ref{fig:OHE}. The Hall coefficient calculated for MRG yielded a value of $R_H = -4.41 \times 10^{-10}$~m$^3$~C$^{-1}$, which corresponds to a carrier concentration $n_e =  1.42\times 10^{22}$\,cm$^{-3}$. In MRG, the Hall effect is predominantly governed by the minority carrier at the Fermi level due to the material's high spin-polarization~\cite{Kurt2014,Fowley2018}.

The planar Hall effect ($R_{xy}^{PHE} $) was evaluated by rotating the magnetic field within the plane of the sample. Figure~\ref{fig:planner} displays the anisotropic magneto-resistance (AMR) and the planar Hall effect (PHE) measured at room temperature in the presence of a magnetic field with a value of $ \SI{1.9}{\tesla} $. The observed PHE is three orders of magnitude smaller than the recorded AHE, thus allowing it to be safely disregarded from equation~\ref{eq:halleffect}. As a result, the primary dominant contribution to the transverse Hall resistance is due to the AHE. Nevertheless, OHE has also been considered in the model to acknowledge its significant contribution, particularly at high applied magnetic fields.

\begin{figure*}[htb]
	\centering
    \begin{minipage}[c]{0.31\linewidth}
    \centering
    \subfloat[\label{fig:OOP_1}]{
    \includegraphics[width=0.65\linewidth]{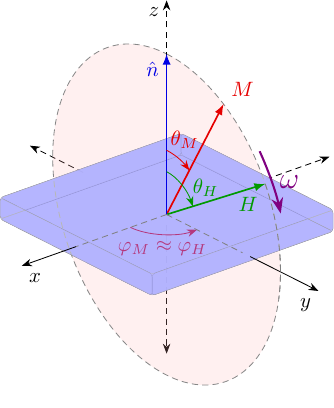}
    }
    \end{minipage}
    \hfil
    \begin{minipage}[c]{0.31\linewidth}
    \centering
    \subfloat[\label{fig:OOP_loop}]{
    \includegraphics[width=\linewidth]{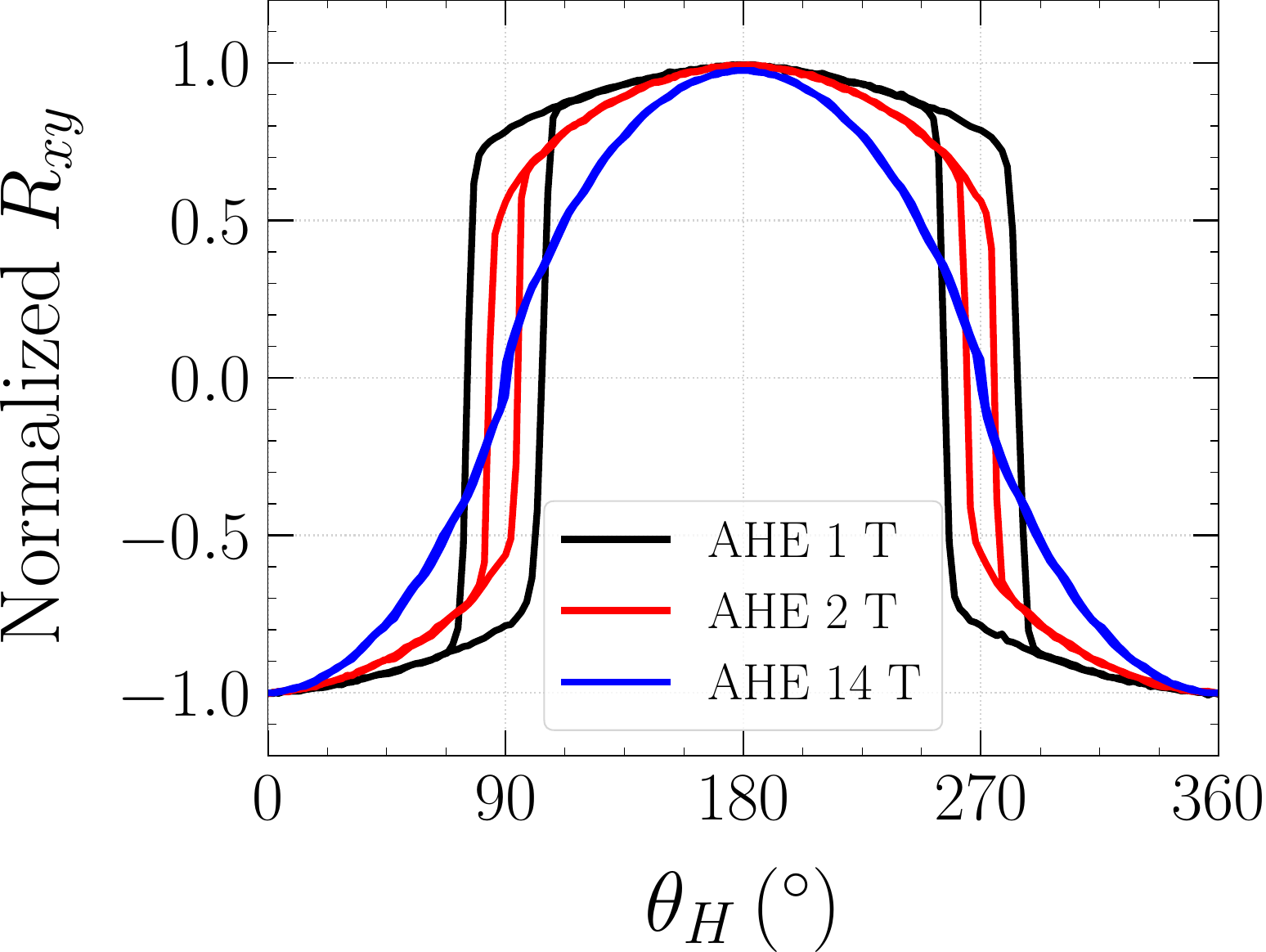}
    }
    \end{minipage}
    \hfil
    \begin{minipage}[c]{0.31\linewidth}
    \centering
    \subfloat[\label{fig:OOP_loop_fit}]{
    \includegraphics[width=\linewidth]{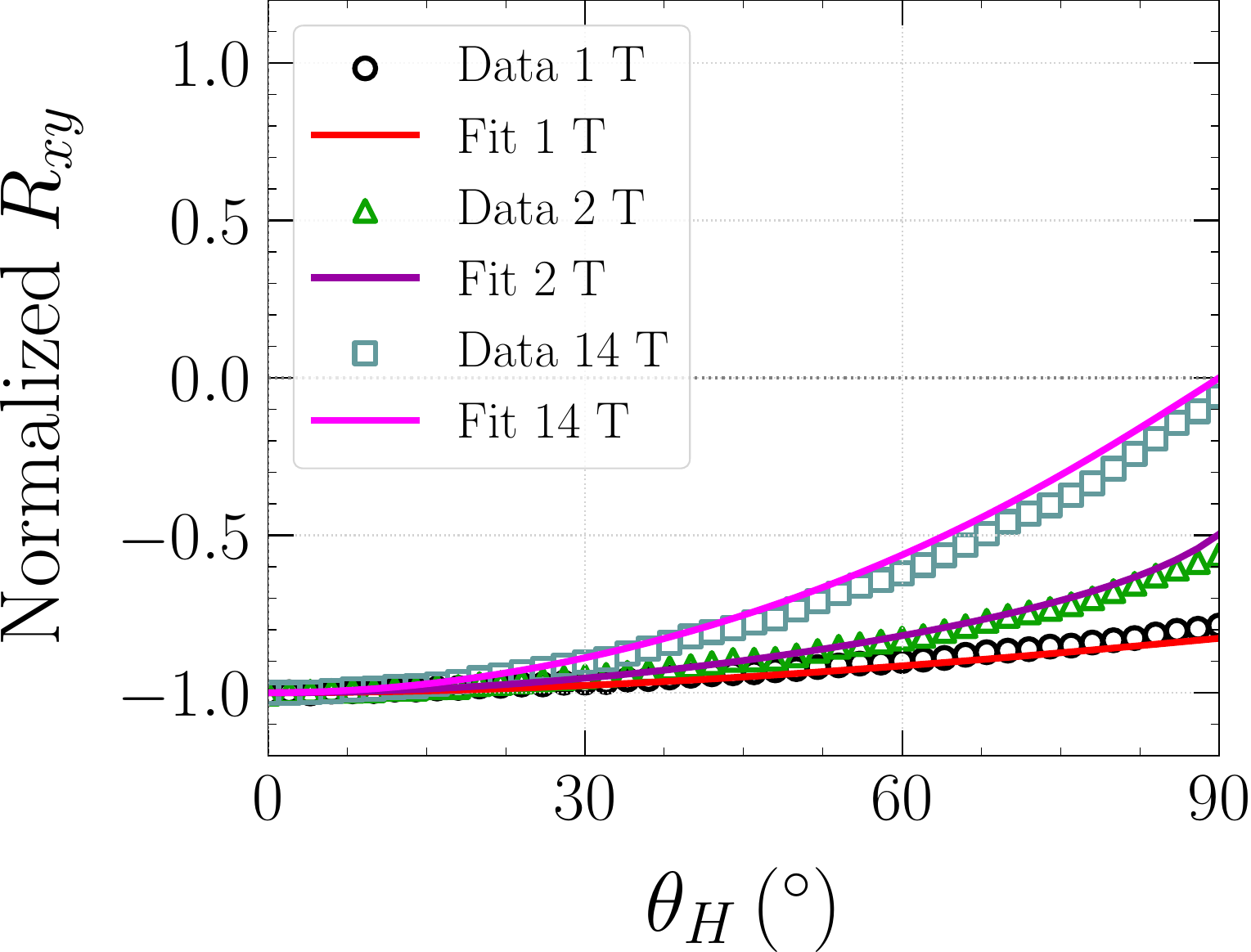}
    }
    \end{minipage}
    \hfil
    \begin{minipage}[c]{0.31\linewidth}
    \centering
    \subfloat[\label{fig:inplane_2}]{
    \includegraphics[width=0.65\linewidth]{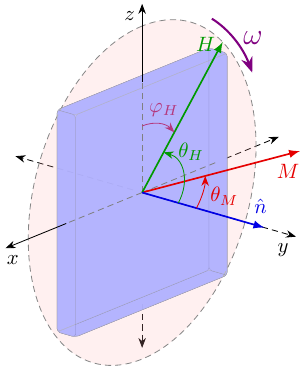}
    }
    \end{minipage}
    \hfil
    \begin{minipage}[c]{0.31\linewidth}
    \centering
    \subfloat[\label{fig:inplane_1T}]{
    \includegraphics[width=\linewidth]{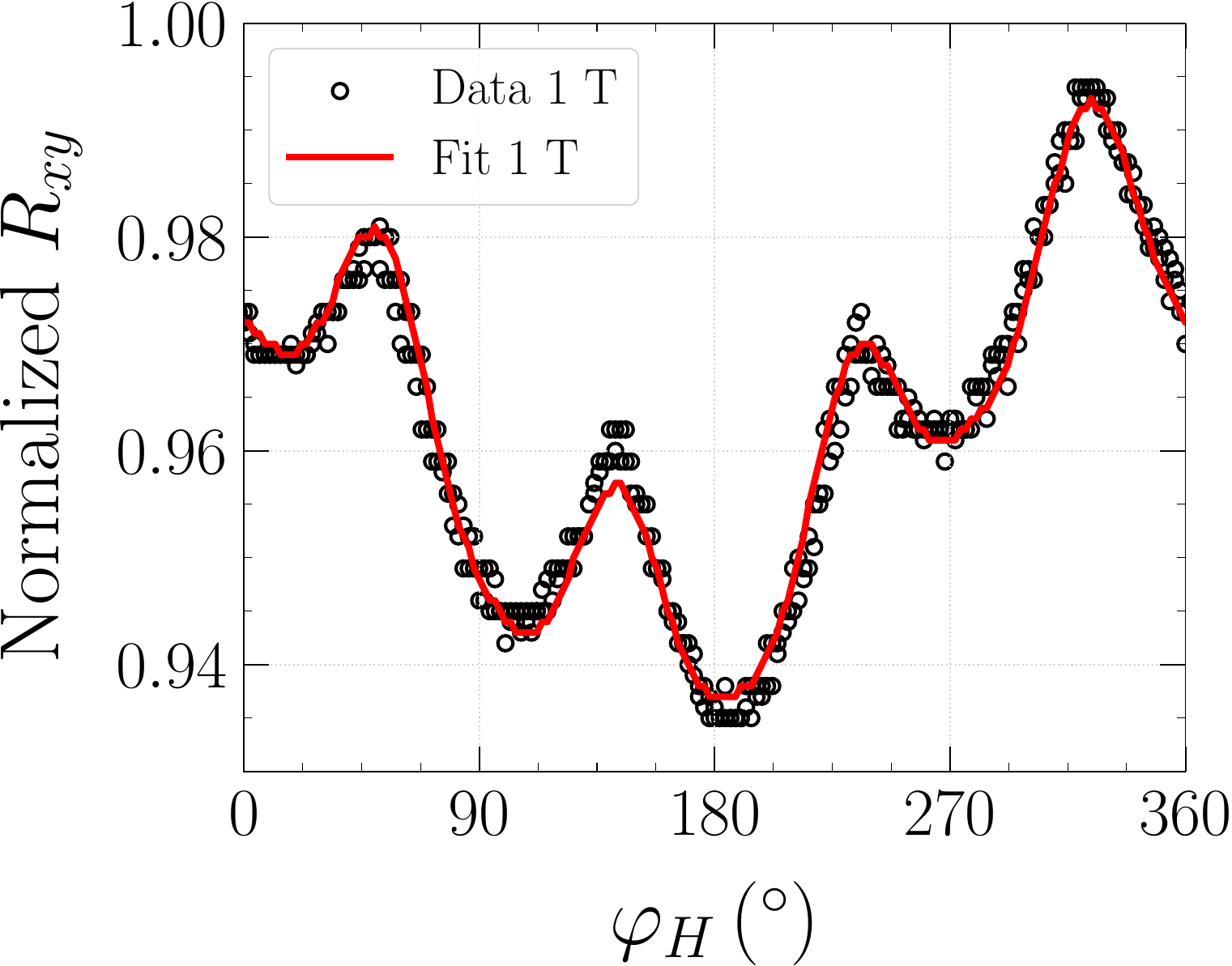}
    }
    \end{minipage}
    \hfil
    \begin{minipage}[c]{0.31\linewidth}
    \centering
    \subfloat[\label{fig:inplane_1p5T}]{
    \includegraphics[width=\linewidth]{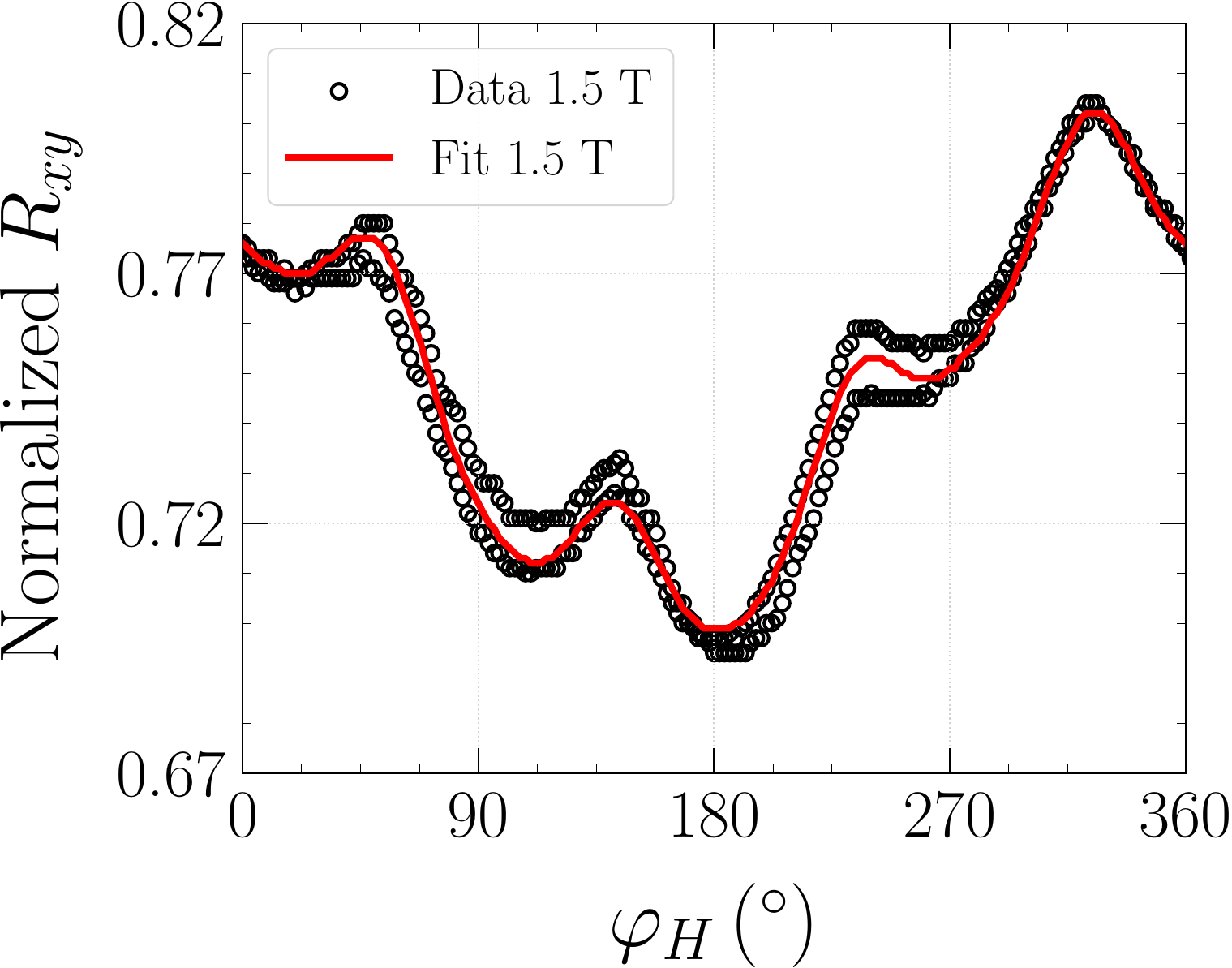}
    }
    \end{minipage}
	\caption{AHE measurement at \SI{300}{\kelvin} in two different geometric arrangements to evaluate the out-of-plane anisotropy constants ($K_1$, $K_2$) and in-plane anisotropy constant ($K_3$) of MRG. (a) Out-of-plane rotational measurement geometry for investigating $K_1$ and $K_2$. In this configuration, a constant applied magnetic field ($H$) is effectively rotated in the $yz$-plane, varying the field angle $\theta_H$ and consequently changing the magnetization angle $\theta_M$, which is recorded using AHE. (b) Rotational AHE data as a function of field angle ($\theta_H$) for constant applied magnetic fields of \SI{1}{\tesla}, \SI{2}{\tesla}, and \SI{14}{\tesla}. Each loop contains two distinct regimes: a hysteretic part attributed to the abrupt switching of magnetization, and a non-hysteretic part due to the coherent rotation of magnetization experiencing net torque. The detailed fits to this data are presented on figure~\ref{fig:OOP_fit}. (c) The recorded non-hysteretic segments of curves (scatter plots) for $\theta_H < 90$$^{\circ}$ are fitted with the balanced torque model (solid lines), given by equation~\ref{eq:GST}. Anisotropy constants $ K_1/M=\SI{0.655}{\tesla} $, $ K_2/M=\SI{0.416}{\tesla} $ are obtained by fitting these AHE curves. (d) In-plane measurement geometry for investigating $K_3$, where the field is effectively rotated within the sample plane (i.e., $\theta_H \approx 90$$^{\circ}$ and $\varphi_H$ is varied), resulting in the modulation of $\theta_M$ as a function of $\varphi_H$. (e) In-plane AHE data (black scattered circles) as a function of field rotation angle ($\varphi_H$) for a field value of \SI{1}{\tesla}. The four troughs and valleys in the curve are associated with the four-fold in-plane anisotropy of MRG. The unequal amplitude of oscillation is attributed to sample offset (sample plane slightly tilted away from the $xz$-plane). Incorporating sample offset in the torque model provides an excellent estimation of in-plane rotational AHE (red solid line), with the estimated $ K_3/M=\SI{0.057}{\tesla} $. (f) In-plane AHE data (black scattered circles) as a function of $\varphi_H$ for an applied field of \SI{1.5}{\tesla}. The estimated curve (red solid line) is obtained for $ K_1/M=\SI{0.655}{\tesla} $, $ K_2/M=\SI{0.416}{\tesla} $, and $ K_3/M=\SI{0.057}{\tesla} $.
    }
	\label{fig:OOP}
\end{figure*}

To investigate the out-of-plane anisotropy constants ($K_1$ and $K_2$), the AHE was conducted in the measurement geometry depicted in figure~\ref{fig:OOP_1}. In this configuration, the sample was rotated in such a way that the applied magnetic field effectively rotated within the $ yz $-plane. Figure~\ref{fig:OOP_loop}  presents the three rotational AHE loops measured at T = 300 K, under constant applied magnetic fields of  $ \SI{1}{\tesla}$, $\SI{2}{\tesla}$, and $\SI{14}{\tesla}$.
The acquired data exhibit two distinct regimes: the first regime showcases a continuous change in the resistance (non-hysteretic segments, for $\theta_H < 90$$^{\circ}$, $\theta_H > 270$$^{\circ}$ and in the vicinity of $\theta_H = 180$$^{\circ}$), attributable to the smooth coherent rotation of the magnetization against the anisotropy field, while the second regime displays an abrupt change in the resistance (hysteretic segments, for $90^{\circ}<\theta_H < 270^{\circ}$) due to the switching of the magnetic moment from out-of-plane to in-plane or vice versa. The non-hysteretic portions of the data were fitted using the torque model. Since the anisotropy constants are independent of the applied external field (at least up to the first order), the non-hysteretic segments of the data should be modelled using common fitting parameters. Figure~\ref{fig:OOP_loop_fit} illustrates the recorded data alongside the corresponding best fits utilizing common anisotropy parameters. The data align well with the model, yielding anisotropy constants of  $ \displaystyle{\frac{K_1}{M}=\SI{0.655}{\tesla}} $  and $ \displaystyle{\frac{K_2}{M}=\SI{0.416}{\tesla}} $, where $ M $  denotes the magnitude of saturation magnetization. The sample's saturation magnetization, obtained from SQUID measurement, is \SI{61}{\kilo\ampere\per\meter}. Consequently, the first and second order out-of-plane anisotropy constants of MRG are 
 $K_1=4.0\times10^4$ J m$^{-3}$ and $K_2=2.54\times10^4$ J m$^{-3}$, respectively. It is crucial to note that in this measurement geometry, the AHE is insensitive to in-plane anisotropy due to the absence of azimuthal rotation of magnetization. The in-plane anisotropy constant ($ K_3 $) can be examined in a measurement geometry where the azimuthal direction of magnetization is varied.

Progressing with the study, the in-plane anisotropy was examined using the AHE in the measurement geometry depicted in figure~\ref{fig:inplane_2}. In this configuration, the sample was rotated in such a way that the applied magnetic field effectively rotated within the plane of the sample ($ xz $-plane). The presence of in-plane anisotropy causes the AHE signal to oscillate as a function of the azimuthal angle ($\varphi_M$) of the magnetization, revealing the four-fold anisotropy of MRG. Figure~\ref{fig:inplane_1T} and \ref{fig:inplane_1p5T} display the scans obtained at \SI{300}{K} when constant magnetic fields of \SI{1}{T} and \SI{1.5}{T} were applied in the plane of the sample, respectively. The unequal amplitude of oscillation arises from a small offset ($\sim$ \SI{6}{\deg}) of the sample from the $ xz $-plane, which subsequently causes sample wobbling during rotation and introduces an additional term – a non-zero normal component – affecting the magnetization vector position $\theta_M$.
The equilibrium position of the magnetization vector, under the influence of the external magnetic field, is numerically obtained using the torque model with a correction for wobbling taken into consideration (the implementation involved employing Rodrigues' rotation formula, which utilizes the appropriate axis of rotation). The extracted value of the in-plane anisotropy constant is $ \displaystyle{\frac{K_3}{M}=\SI{0.057}{\tesla}} $, or $K_3=3.48\times10^3$ J m$^{-3}$, which is an order of magnitude smaller than the out-of-plane anisotropy constants $K_1$ and $K_2$. It is noteworthy that an increase in field strength results in a more pronounced hysteresis of AHE in the in-plane configuration (comparing figure~\ref{fig:inplane_1T} and \ref{fig:inplane_1p5T}). This phenomenon occurs because, at sufficiently high magnetic fields, the sample wobbling leads to partial switching of magnetic moments. Modeling such complex data, where both coherent rotation and magnetization switching take place, can be accomplished by combining both the Preisach and torque models.

\subsection{Combined Preisach and torque (CPT) model}
\label{sec:combined_model}
\subsubsection{In-plane field hysteresis loop}\label{sec:inplane_1}
\begin{figure*}[htb]
	\centering
    \begin{minipage}[c]{0.40\linewidth}
    \centering
    \subfloat[\label{fig:inplane_3}]{
    \includegraphics[width=0.68\linewidth]{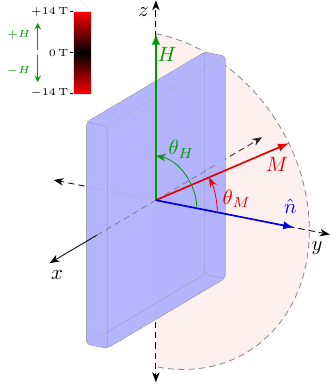}
    }
    \end{minipage}
    \hfil
    \begin{minipage}[c]{0.40\linewidth}
    \centering
    \subfloat[\label{fig:inplane_FS1}]{
    \includegraphics[width=\linewidth]{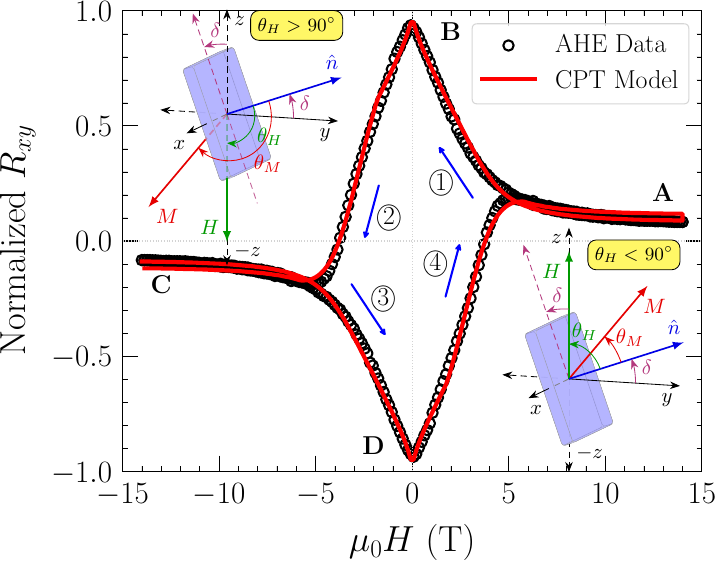}
	}
    \end{minipage}
    \begin{minipage}[c]{0.40\linewidth}
    \centering
    \subfloat[\label{fig:inplane_hysteron}]{
    \includegraphics[width=\linewidth]{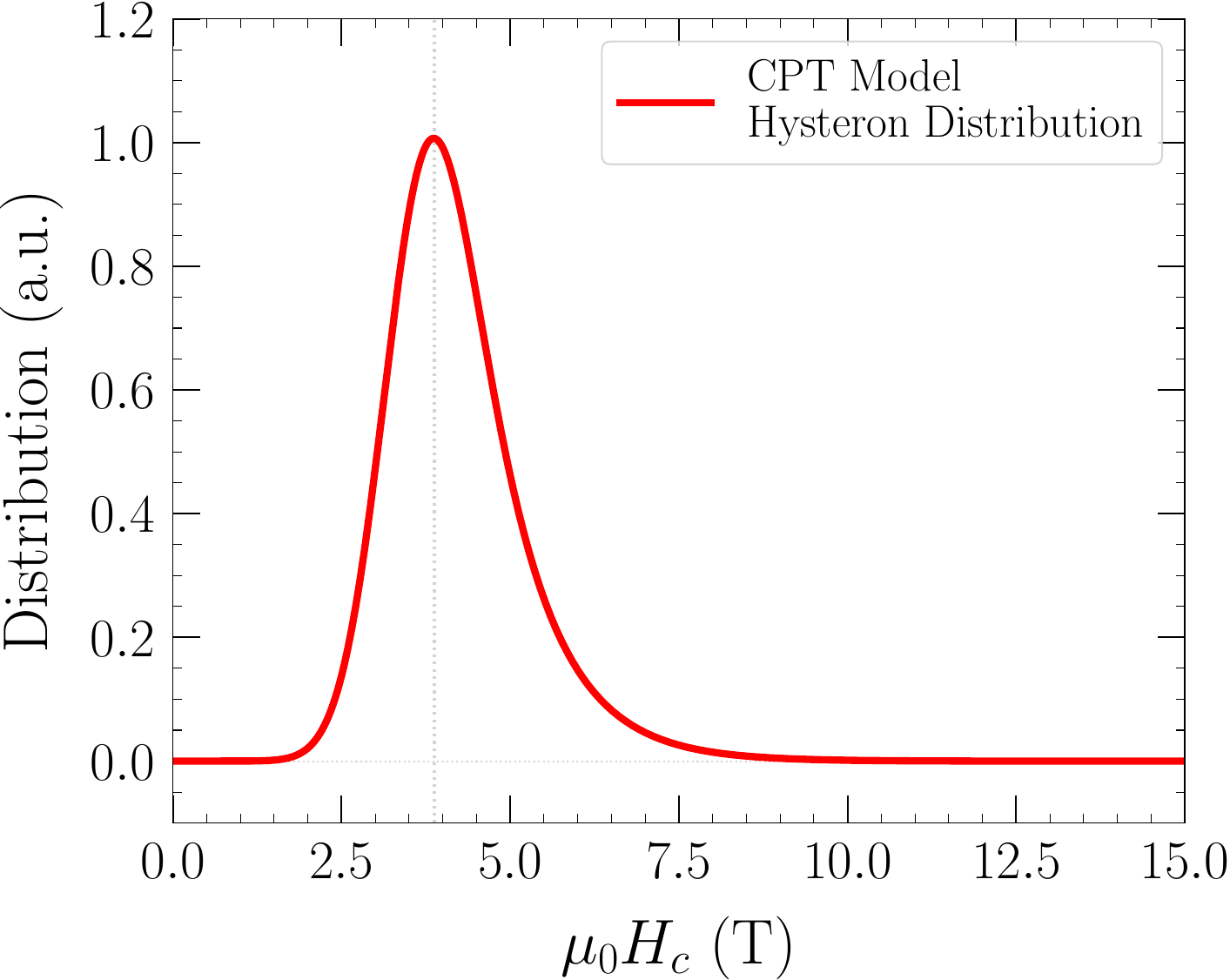}
    }
    \end{minipage}
    \hfil
    \begin{minipage}[c]{0.40\linewidth}
    \centering
    \subfloat[\label{fig:inplanefs2}]{
    \includegraphics[width=\linewidth]{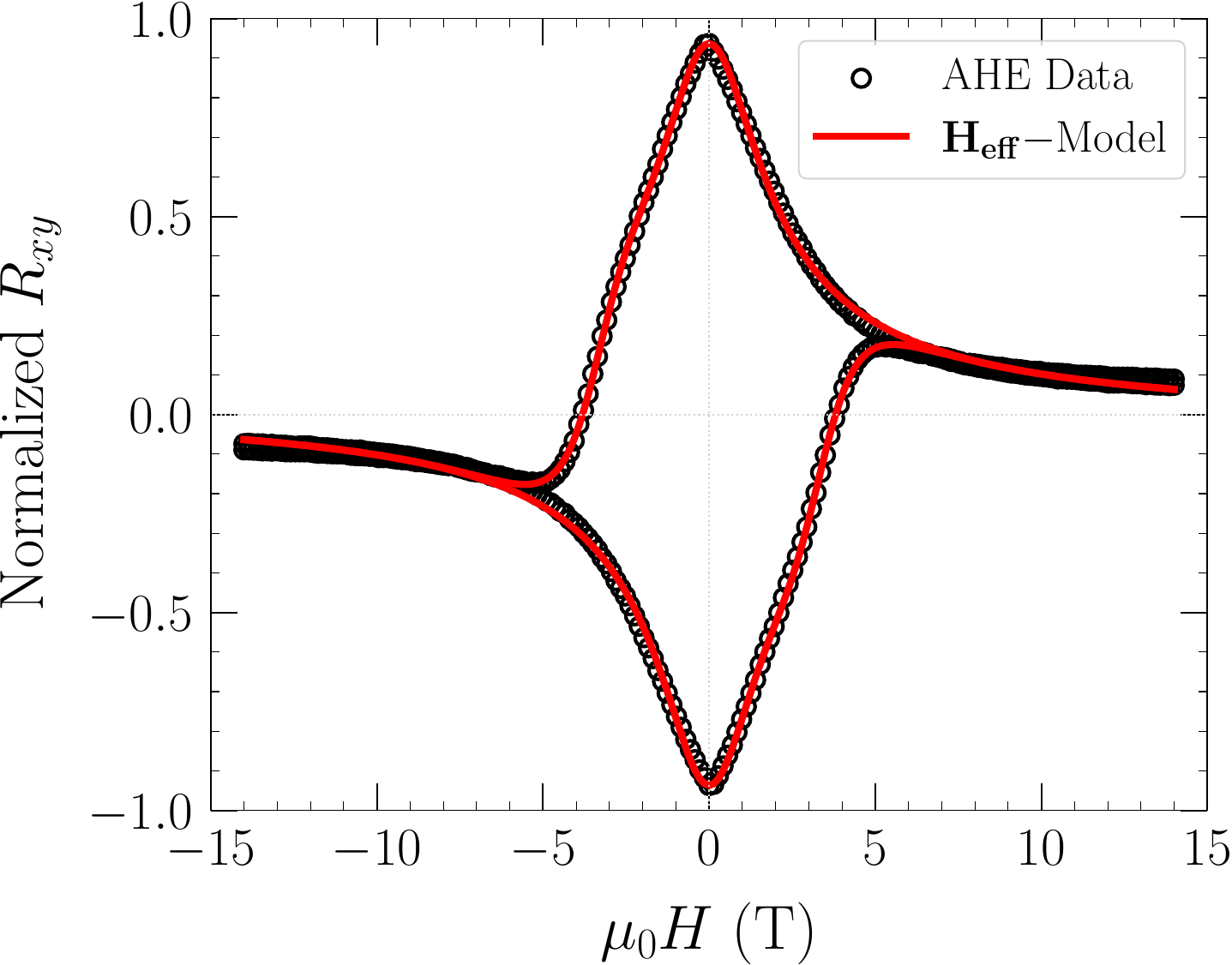}
    }
    \end{minipage}
	\caption{In-plane field loop study for the combined hysteresis and torque model at \SI{300}{\kelvin}. (a) Measurement geometry of in-plane field loop study, where the magnetic field is swept between \SI{\pm14}{\tesla} in the sample plane ($\theta_H \approx 90\si{\degree}$), along the $z$-axis. (b) Resultant experimentally obtained AHE curve (black open circles). An unavoidable sample offset ($\delta$) with respect to the field axis ($z$) leads to this intricate magnetization dynamics, where the magnetization direction switches sign when the normal component of the field exceeds the sample's coercivity. The insets illustrate the geometrical alignment of magnetization and applied magnetic field when the field angle ($\theta_H$) forms an acute angle with the sample's normal direction $\hat{n}$ (right inset) and an obtuse angle with $\hat{n}$ (left inset). Paths \textcircled{\raisebox{-0.9pt}{1}} and \textcircled{\raisebox{-0.9pt}{3}} represent portions of the curve where magnetization experiences coherent rotation, while paths \textcircled{\raisebox{-0.9pt}{2}} and \textcircled{\raisebox{-0.9pt}{4}} display the behavior when magnetization dynamics is dominated by switching events. The estimated AHE (red line) within the combined Preisach and torque (CPT) model shows excellent agreement with the data. (c) Coercivity distribution (hysteron distribution) derived from the CPT model. This distribution curve also exhibits strong resemblance to the hysteron distribution curve obtained when the magnetic field is applied perpendicular to the sample (as shown in figure~\ref{fig:FORC_Hc}), with the field axis scaled according to equation~\ref{eq:coercive}. (d) AHE hysteresis data (black circles) modeled using an effective out-of-plane anisotropy approximation (red line). Under this approximation, the torque model is simplified with a single out-of-plane anisotropy field ($\mathbf{H_{eff}}$). By directly incorporating a predetermined hysteron distribution into the model, the complexity of the model is greatly reduced, with the added benefit of having a minimal number of free parameters. 
    }
	\label{fig:inplane_FS}
\end{figure*}

The intricate quasi-static magnetization dynamics can be elucidated by combining the torque and Preisach models, wherein the magnetic moment exhibits both coherent rotation and abrupt switching events under the influence of suitable stimuli. One such mixed behavior can also be observed when a high magnetic field is swept within the plane of the sample.
To examine such dynamics, AHE was measured in the measurement geometry depicted in figure~\ref{fig:inplane_3}. In this setup, the magnetic field was swept within the plane of the sample (along the $z$-axis) from $\pm$\SI{14}{\tesla}. The recorded AHE signal exhibits a combination of hysteretic and rotational behavior, as shown in figure~\ref{fig:inplane_FS1}. Due to an  unavoidable  minor offset ($\delta$) of the sample while mounting it on the rotary stage of the PPMS, the applied magnetic field does not lie exactly within the plane of the sample (see the inset of figure~\ref{fig:inplane_FS1}).
An offset of approximately  $\theta\sim$\,\SI{6}{\deg} was present, the exact offset value can be determined by fitting the AHE data. As a result, the AHE signal has two distinct regimes: (i) when the magnetic field (H) forms an acute angle with the normal to the sample ($\theta_H< \,$\SI{90}{\deg}), a coherent rotation of the magnetization vector is observed; and (ii) when the magnetic field makes an obtuse angle with the normal ($\theta_H> $\,\SI{90}{\deg}), the switching of the magnetic moment occurs when the field projection surpasses the coercive field. It should be noted that the coercive field in this scenario is scaled according to equation~\ref{eq:coercive}:
\begin{equation}\label{eq:coercive}
	H_c=\frac{H_{c_n}}{\cos(\theta_H)}\, ,
\end{equation}
here, $H_{c_n}$ represents the coercivity of the MRG sample when the magnetic field is applied along the direction normal to the sample plane ($\theta_H=0$).

A comprehensive trajectory of magnetization under the influence of the applied magnetic field is meticulously demonstrated in figure~\ref{fig:inplane_FS1}. This representation encompasses the combined behaviour of the rotation of magnetic moments and their corresponding switching events, which can be effectively described through equation~\ref{eq:combine};
\begin{equation}\label{eq:combine}
	R_{xy}= R_{xy}^{TM}\cdot R_{xy}^{PM}
\end{equation}
where,  $R_{xy}^{TM}$ and $R_{xy}^{PM}$ denote the respective contributions arising from the torque model (as explicated in equation~\ref{eq:GST} ) and the Preisach model (as detailed in equation~\ref{eq:hys_mod}). To determine the solution for equation~\ref{eq:combine}, the Levenberg-Marquardt algorithm was employed, while maintaining the anisotropy constants $K_1/M$ at a fixed value of \SI{0.655}{\tesla} and $K_2/M$ at \SI{0.416}{\tesla}. The resultant estimated curve, in conjunction with the data, is displayed in figure~\ref{fig:inplane_FS1}, which reveals an exceptional congruence with the gathered data.
Furthermore, figure~\ref{fig:inplane_hysteron} depicts the estimated hysteron distribution for the corresponding AHE curve, with the central point of distribution ($H_{c_0}$) determined to be \SI{3.88}{\tesla}. Notably, this distribution curve also exhibits a strong resemblance to the hysteron distribution curve obtained when the magnetic field is applied perpendicular to the sample (as shown in figure~\ref{fig:Hy_300K}), with the field axis scaled according to equation~\ref{eq:coercive}. This consistency highlights the robustness of the analysis and further validates the effectiveness of the model.

Though the combined Preisach and torque (CPT) model, as delineated by equation~\ref{eq:combine}, effectively predicts the intricate magnetization dynamics, implementing this equation to depict complex quasi-static magnetic dynamics presents a considerable computational challenge due to the multiparameter nature of the equation. Nevertheless, the awareness that the Preisach distribution for a specific temperature can be independently determined through a pure switching event (out-of-plane hysteresis curve) allows for further simplification of the combined model. This is achieved by further considering an effective out-of-plane anisotropy field ($\mathbf{H_{eff}}$) to resolve the pure torque model component. The rationale for utilizing an effective anisotropy field stems from the fact that MRG exhibits substantial and dominating out-of-plane anisotropy, which is also evident from the steep square hysteresis loop observed when the field is swept perpendicular to the sample (figure~\ref{fig:OHE}). Under this approximation, the equilibrium position of magnetization ($\mathbf{M}$) can be attained by counterbalancing the torques acting upon it, (equation~\ref{eq:torque_2} ) 

\begin{equation}\label{eq:torque_2}
	\mathbf{M}\times\mu_0\mathbf{H_{eff}} = \mathbf{M}\times\mu_0\mathbf{H}\, ,
\end{equation}

where, $\mathbf{H_{eff}}$ and $\mathbf{H}$  represent the effective out-of-plane anisotropy field and the applied external magnetic field, respectively. In figure~\ref{fig:inplanefs2}, the data and corresponding fit are presented, which utilize the $\mathbf{H_{eff}}$  model with a single free parameter ($H_{eff}$ ) and the Preisach model that has been determined previously. The calculated $H_{eff}$  value from the fitting is \SI{1.46}{\tesla}. The model captures all details of the AHE, thereby validating the proposed approximation. It is important to note that by comparing figures~\ref{fig:inplane_FS1} and ~\ref{fig:inplanefs2}, the distinction between the two models can be discerned. At high magnetic field values ($|\mu_0H| \geq \SI{5}{\tesla}$), the effective anisotropy field model slightly deviates from the data and does not accurately capture the curvature of the data as effectively as the complete model (figure~\ref{fig:inplane_FS1}). This is due to the model's assumption of a unique fixed ${H_{eff}}$ value for all $\mathbf{M}$ orientations. In contrast, the magnitude of ${H_{eff}}$ for a tetragonal crystal system relies on the magnetization direction, and its magnitude typically decreases as $\mathbf{M}$ deviates from the out-of-plane direction (easy-axis). Consequently, a ${H_{eff}(\theta_M)}$ is necessary to capture the data in greater detail for all possible magnetic field values.
Nonetheless, it is adequate to assume that a single fixed  ${H_{eff}}$ performs remarkably well, at least up to the magnetic field strength employed in this study ($ \SI{14}{\tesla} \leq |\mu_0H|$). This approximation offers a significant advantage in describing complex magnetization dynamics by substantially reducing the number of free parameters in the model.

\subsubsection{Out-of-plane rotational hysteresis loop}
\begin{figure*}[htb]
	\centering
    \begin{minipage}[c]{0.40\linewidth}
    \centering
    \subfloat[\label{fig:OOP_2}]{
    \includegraphics[width=0.62\linewidth]{OOP.pdf}
    }
    \end{minipage}
    \hfil
    \begin{minipage}[c]{0.40\linewidth}
    \centering
    \subfloat[\label{fig:OOP_1T}]{
    \includegraphics[width=\linewidth]{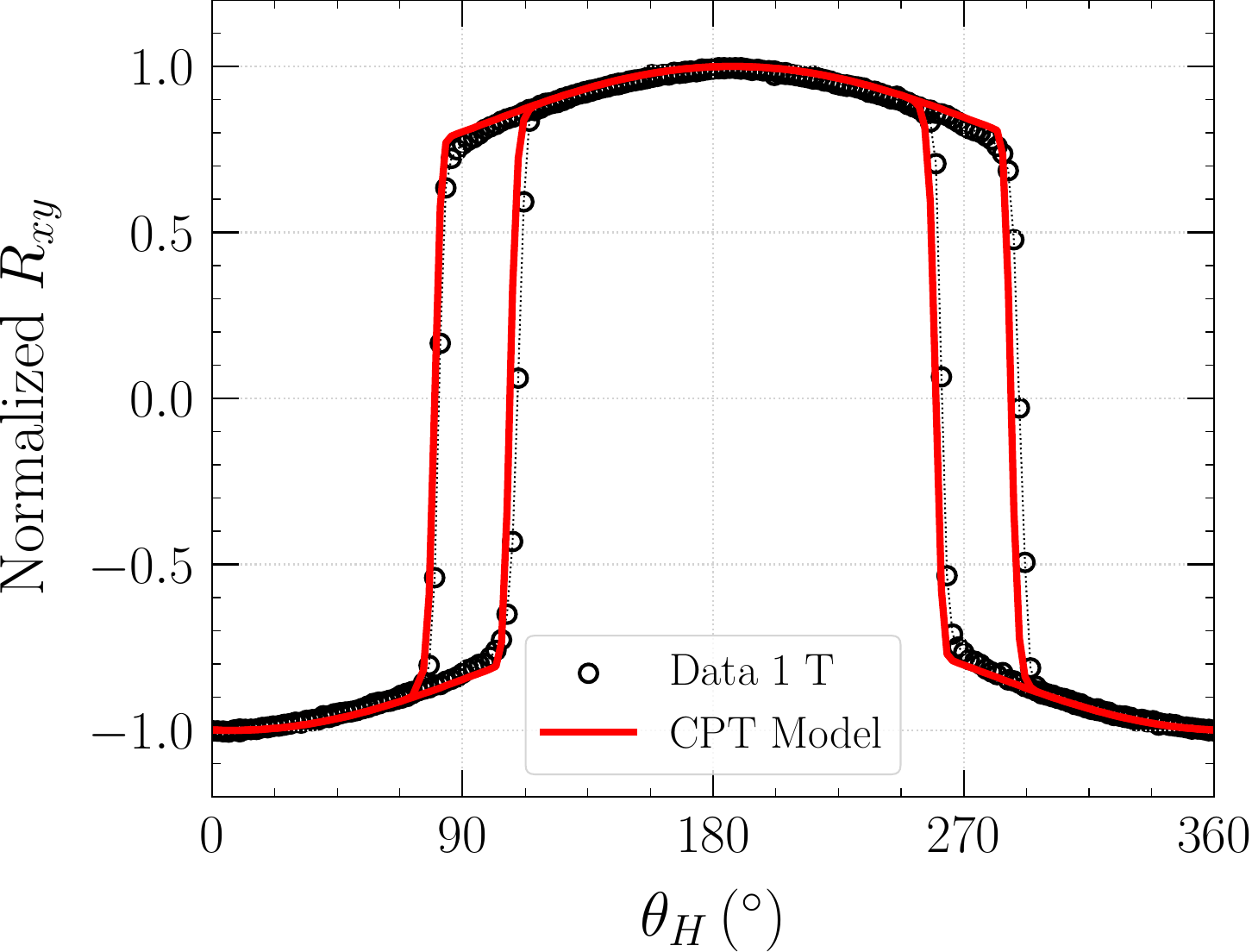}
    }
    \end{minipage}
    \begin{minipage}[c]{0.40\linewidth}
    \centering
    \subfloat[\label{fig:OOP_2T}]{
    \includegraphics[width=\linewidth]{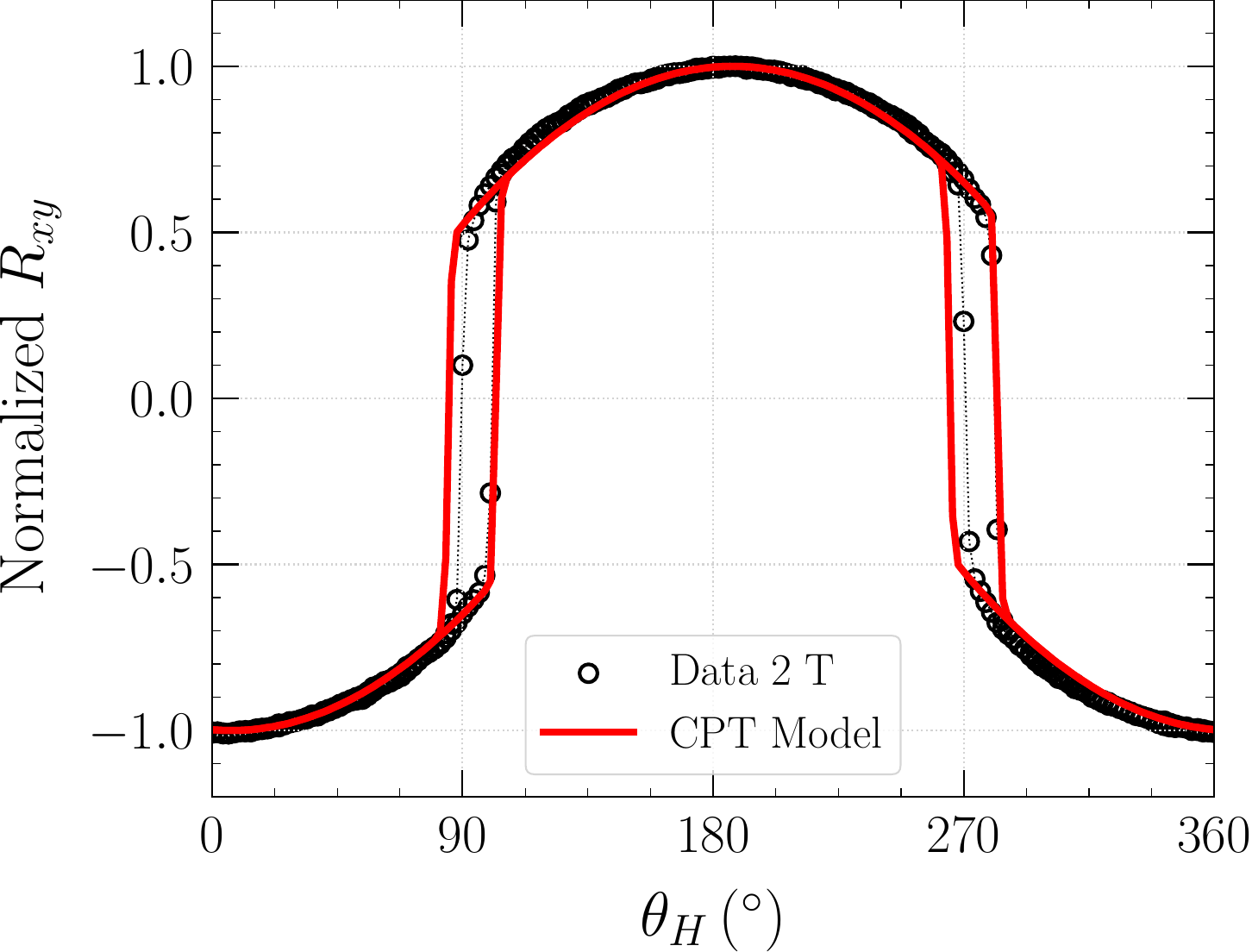}
    }
    \end{minipage}
    \hfil
    \begin{minipage}[c]{0.40\linewidth}
    \centering
    \subfloat[\label{fig:OOP_14}]{
    \includegraphics[width=\linewidth]{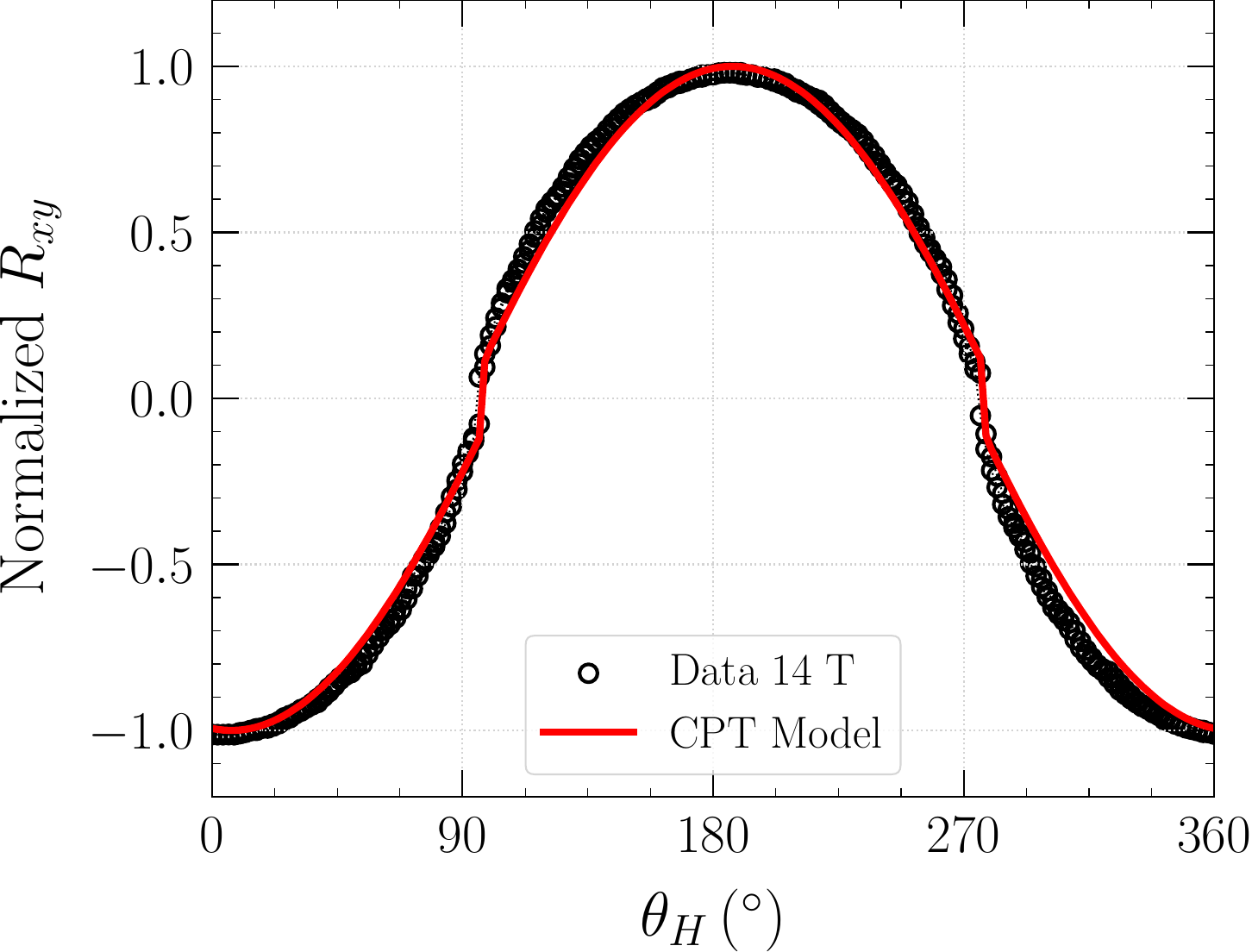}
    }
    \end{minipage}
	\caption{Investigation of out-of-plane rotational hysteresis loop within the CPT model. (a) Measurement geometry where a constant magnetic field is effectively rotated in the $yz$-plane. As a result, the magnetization angle ($\theta_M$) changes within the $yz$-plane as the field angle ($\theta_H$) varies. Note that this set of data have already been presented in figure~\ref{fig:OOP_loop}. AHE loops as a function of $\theta_H$ for the applied field values of \SI{1}{\tesla}, \SI{2}{\tesla} and \SI{14}{\tesla} are shown in (b), (c), and (d) respectively. The CPT model under the effective anisotropy ($\mathbf{H_{eff}}$) approximation accurately captures the behavior of the data at each field (red line). All parameters, such as the effective out-of-plane anisotropy field ($H_{eff}$) and the coefficients of the hysteron distribution ($H_{c_0},\Gamma~ \text{and}\,\tau$), remain constant in the current CPT fitting approach. These parameters were determined beforehand by fitting other AHE curves, as explained in the preceding sections.
    }
	\label{fig:OOP_fit}
\end{figure*}

The efficacy of the CPT model is further substantiated by applying it to magnetization dynamics derived from out-of-plane rotational hysteresis curves, as illustrated in figure~\ref{fig:OOP_fit}. In this experiment, AHE curves were acquired utilizing the measurement geometry shown in figure~\ref{fig:OOP_2}. This setup involves rotating a constant applied magnetic field within the $yz$-plane, causing the equilibrium position of the magnetic moment ($\theta_M$) to reside within the same plane. A collection of AHE data, recorded at applied magnetic field values of \SI{1}{\tesla}, \SI{2}{\tesla} and \SI{14}{\tesla}, and their corresponding CPT fits, are displayed in figures~\ref{fig:OOP_1T}, ~\ref{fig:OOP_2T}, and ~\ref{fig:OOP_14}, respectively. The CPT model well-describes the dataset across all applied magnetic fields.
It is important to note that as the magnitude of the applied magnetic field escalates, the hysteretic contribution to the AHE starts to decrease relative to the non-hysteretic contribution, resulting in a reduced hysteretic width. In cases where the field strength reaches exceptionally high levels, the Zeeman term prevails over the anisotropy term, thereby causing the magnetization to effectively align with the magnetic field direction. Consequently, at a \SI{14}{\tesla} field, the hysteresis width has virtually disappeared (figure~\ref{fig:OOP_14}).
In the context of the current CPT fitting approach, all parameters, including the effective out-of-plane anisotropy field ($H_{eff}$) and the coefficients of hysteron distribution ($H_{c_0},\Gamma~ \text{and}\,\tau$), are maintained as constant values. These parameters have been previously determined through the fitting of other AHE curves, as elaborated upon in the preceding sections. As a result, the derived fitting curve successfully captures both the hysteretic and non-hysteretic aspects of the AHE curve with remarkable precision, for both low ($\mu_0H = \SI{1}{\tesla}$) and high applied magnetic fields ($\mu_0H = \SI{14}{\tesla}$), while virtually eliminating the need for free parameters.

\section{Conclusion}
\label{sec:conclusion}
In this work, we have developed a comprehensive methodology for determining the various magnetic anisotropy constants of low-moment MRG thin films. To achieve this, we initially investigated hysteretic phenomena using the Preisach model, also known as the hysteron model. The applicability of the Preisach model was subsequently experimentally verified through the implementation of the first-order reversal curves (FORC) method, which enabled us to identify the unique hysteron distribution of the sample under investigation. The FORC method provided crucial insights, specifically highlighting the absence of long-range magnetic interactions within the hysterons, which allowed for the utilization of the macrospin model (Stoner-Wohlfarth model) to describe the quasi-static magnetization dynamics of MRG. Furthermore, the Preisach model confirmed that MRG samples exhibit relatively weak variations in magnetic viscosity with temperature, signifying the presence of a frozen domain structure.
To determine the anisotropy constants of the MRG samples, we employed a detailed torque model within the macrospin approximation framework. Anomalous Hall effect (AHE) measurements were carried out in various suitable geometries, which facilitated the deduction of out-of-plane anisotropy constants $K_1=4.0\times10^4$ J m$^{-3}$ ($K_1/M=0.655$\,T) and $K_2=2.54\times10^4$ J m$^{-3}$ ($K_2/M=0.416$\,T), and an in-plane anisotropy constant $K_3=3.48\times10^3$ J m$^{-3}$ ($K_3/M=0.057$\,T) through data fitting with the torque model.
Additionally, we successfully investigated more complex quasi-static magnetization dynamics, characterized by the combination of hysteretic and non-hysteretic components in AHE, using a combined Preisach and torque (CPT) model with virtually no free parameters. Our study demonstrates the efficacy of this methodology not only in determining the magnetic anisotropy of low moment magnetic samples (MRG), but also in explaining other complex magnetization dynamics within a unified model.
The proposed method can be readily extended to other magnetic systems that lack hysteronic interactions, exhibit narrow hysteron distributions, and display frozen-domain behaviour. This comprehensive approach will undoubtedly prove valuable in studying both linear and non-linear quasi-static magnetization dynamics of MRG in external fields and/or current-induced effective fields resulting from spin-orbit torque/spin-transfer torque.
\section*{Acknowledgements}
A.J., S.L., G.P., K.R., J.M.D.C and P.S. acknowledge funding from TRANSPIRE FET Open H2020 and SFI, AMBER and MANIAC programmes.

\bibliography{mybibliography}

\end{document}